\documentclass[aps,prd,amssymb,amsmath,nofootinbib]{revtex4}
\usepackage{latexsym}
\usepackage{graphics}
\usepackage{graphicx}
\usepackage{feynmf} 
\usepackage{color}
\newcommand{\la}{\langle}
\newcommand{\ra}{\rangle}
\newcommand{\ar}{\arrowvert}
\newcommand{\be}{\begin{equation}}
\newcommand{\ee}{\end{equation}}
\newcommand{\ba}{\begin{eqnarray}}
\newcommand{\ea}{\end{eqnarray}}
\newcommand{\mc}[1]{\mathcal{#1}}
\newcommand{\pa}{\partial}
\newcommand{\tr}[1]{\textrm{#1}}
\newcommand{\bw}{\begin{widetext}}
\newcommand{\ew}{\end{widetext}}
\newcommand{\ov}{\overline}

\begin{document}
\title{Minimum of $\eta/s$ and the phase transition \\
of the Linear Sigma Model in the large--N limit}
\author{Antonio Dobado, Felipe J. Llanes-Estrada and Juan M. Torres-Rincon}
\affiliation{Departamento de F\'isica Te\'orica I, Universidad Complutense de Madrid, 28040 Madrid, Spain}
\begin{abstract}
  We reexamine the possibility of employing the viscosity over entropy density ratio as a diagnostic tool to identify a phase transition in hadron physics to the strongly coupled quark--gluon plasma and other circumstances where direct measurement of the order parameter or the free energy may be difficult.
  It has been conjectured that the minimum of $\eta/s$ does indeed occur at 
the phase transition. We now make a careful assessment in a controlled theoretical framework, the Linear Sigma Model at large--$N$, and indeed find that the minimum of $\eta/s$ occurs near the second-order phase transition of the model due to the rapid variation of the order parameter (here the $\sigma$ vacuum expectation value) at a temperature slightly smaller than the critical one.
\end{abstract}
\maketitle
\section{Introduction}
The viscosity over entropy density ratio $\eta/s$ has recently received much attention due to the conjectured universal bound \cite{Kovtun:2004de} $\eta/s \ge \frac{1}{4\pi}$
for fluids describable by any quantum field theory. No experimental exception to this rule has been found to date, and many have been examined, such as ordinary gases and liquids like argon \cite{Dobado:2008vt}, undoped graphene \cite{mueller-2009}, fermions near the unitarity limit \cite{schaefer-2007-76, Turlapovetal}, and more interestingly for our purposes, the hadron phases formed after Relativistic Heavy Ion Collisions \cite{Csernai:2006zz, Gavin:2006xd,Collaboration:2009yd,Demir:2008tr}.
The existence of this bound can be inferred from dimensional arguments  for a quantum fluid alone~\cite{Danielewicz:1984ww}, but the $1/4\pi$ numerical factor has only been obtained in the context of conformal field theories with a gravity dual. In \cite{Buchel:2008vz,Sinha:2009ev} some possible violations to this bound are proposed.

Important insight came from \cite{Csernai:2006zz} as it was further conjectured that the minimum of $\eta/s$ for several ordinary fluids, but also for hadron matter, coincides with the phase transition. We \cite{Dobado:2008vt, Dobado:2008jr} and others \cite{Chen:2007jq, Sasaki:2008um} have gathered further empirical and theoretical evidence in favor of this concept.

Its use is quite evident for the experimental program of FAIR facility (for example the Compressed Baryonic Matter experiment) or RHIC if run at a lower energy. The idea is simply to use the minimum of $\eta/s$, that might be accessible by studying the momentum distribution of final state pions and other particles, through elliptic flow \cite{Gavin:2006xd, Teaney:2003kp} for example, as a diagnostic 
to locate the phase transition and possibly the critical point in QCD.

\begin{figure}
\includegraphics[height=2.2in]{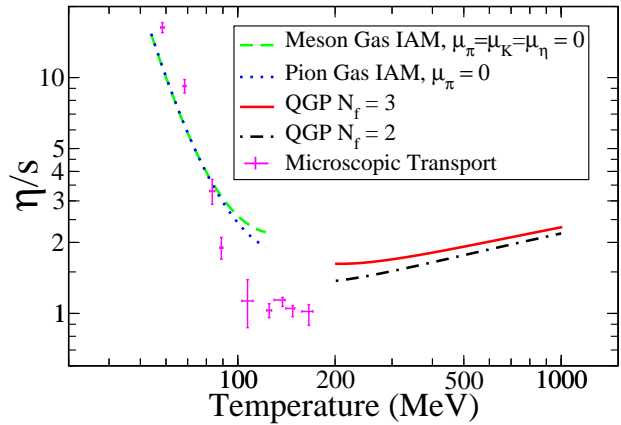}
\caption{The phase transition between a hadron phase (left curves) and a strongly coupled quark--gluon plasma (right curves) points out to a minimum or a discontinuity depending on the order of the transition. At $\mu_B \sim 0$ a soft minimum is expected as hinted in our prediction for $\eta/s$ in the hadron side. The QGP estimations are based on the calculations in \cite{Csernai:2006zz,Arnold:2003zc}, implementing the formulae $\eta/s =5.119/\left[ g^4 \ln \left( 2.414 \ g^{-1} \right) \right] $ for $N_f=3$ and $\eta/s=5.328/ \left[ g^4 \ln \left( 2.558 \ g^{-1}\right) \right]$ for $N_f=2$, with $g$ as a function of temperature given in \cite{Csernai:2006zz}. Finally, the calculation of Demir and Bass with a microscopic transport model is also given \cite{Demir:2008tr}.}
\label{qgpminimum}
\end{figure}

For small baryon chemical potential, and based on lattice QCD calculations, one expects a crossover. Therefore the ratio $\eta/s$ would show a soft minimum when the quark-gluon plasma is about to appear. Our theoretical estimate of $\eta/s$ in this case is shown in Fig.~\ref{qgpminimum}, an update of that in \cite{Dobado:2008vt}. The hadron side (mainly containing pions but also kaons, $\eta$ mesons and their dynamical, elastic low-energy resonances in minor concentration) is calculated by means of the Boltzmann-Uehling-Uhlenbeck (BUU) transport equation and using Chiral Perturbation Theory for the low-energy interactions with the Inverse Amplitude Method for the meson amplitude unitarization. We have not included a finite nucleon density in our work, but this has been estimated independently in Ref.~\cite{Noronha}.

For the quark--gluon plasma side we took the perturbative estimate of \cite{Csernai:2006zz,Arnold:2003zc} for massless quarks with $N_f=2$ and $N_f=3$. In \cite{Dobado:2008vt} we showed that the corrections in this side due to the finite quark masses are very small by using some results in \cite{Aarts:2005vc}.

In this contribution we are going to sharpen the statement somewhat, and show quite clearly that the minimum of $\eta/s$ occurs not exactly at the phase transition, but somewhat before, where the condensate has rapid variation. 

Computation of transport coefficients in a pion gas requires 
knowledge of the $\pi\pi$--scattering amplitude. Although in our recent work \cite{Dobado:2008vt} we have employed Chiral Perturbation Theory and Unitarized Chiral Perturbation Theory, that is the correct low-energy limit of Quantum Chromodynamics, the phase transition is not accessible from it, and one needs to resort to a separate theoretical approach for higher energies (as in Fig.~\ref{qgpminimum}). 
This however would obscure our purpose of showing the correlation between the minimum of $\eta/s$ and the phase transition.

Since some level of modeling is necessary anyway, we adopt from the start
of this article the Linear Sigma Model (L$\sigma$M) in the limit of large--$N$, that although less precise to provide us the scattering amplitudes necessary in the computation of transport coefficients (see Appendix \ref{pipiappendix}), it allows us to control the effective potential and phase transition. Thus, the different phases of the model can be obtained --as desired-- from the same partition function.

An initial study in this direction is \cite{Chen:2007jq}. However, we feel that we can improve and better motivate the characterization of the phase diagram and the location of the critical temperature in the system without resorting to arguments of ``naturalness'', through some additional mathematical effort. In this article we will therefore describe a more exhaustive and consistent study for both the phase diagram of the model and the calculation of $\eta/s$.

The rest of the paper is organized as follows. First we present minimum discussion on the L$\sigma$M effective potential in Sec.~\ref{sec:effV}, but all mathematical detail is relegated to Appendix \ref{app:effVmath}.
Section~\ref{sec:KSS} is a brief indication of how the viscosity over entropy ratio is obtained in the L$\sigma$M, with necessary detail in Appendices
\ref{pipiappendix}, \ref{kineticvisco} and \ref{kubovisco}. 
Our main point, the connection between the fast--changing condensate and the minimum of $\eta/s$, is presented in Sec.~\ref{sec:minimum}.
We also comment on the possibility of identifying the critical exponent of the viscosity near the critical end--point in the phase diagram in Sec.~\ref{sec:critical}, and expose the behavior of $\eta/s$ in helium-4 at low temperature.
Finally, Sec.~\ref{sec:conclusions} summarizes our main results and looks ahead to what interesting future work there might be. Appendices~\ref{app:density} and \ref{app:polemass} provide additional technical aspects supporting our results, such as why the minimum of $\eta/s$ remains below $T_c$, and the applicability of the transport equation.

\section{Effective potential and phase transition\label{sec:effV}}

In this Section we will obtain the phase diagram in order to characterize the phase transition 
temperature of the L$\sigma$M at large--$N$. The physics of the L$\sigma$M at large--$N$ has been extensively described at zero-temperature in \cite{Coleman:1974jh, Dobado:1994fd, Dobado:1997jx} (for both zero and nonzero pion mass), at finite temperature in \cite{Petropoulos:2004bt} and the transport coefficients in \cite{Aarts:2003bk, Aarts:2004sd}, for instance.

The well-known renormalizable Lagrangian density for an $N+1$ multiplet $\Phi$ is
\be \label{lagrangian} \mathcal{L} [\Phi, \pa_{\mu} \Phi] = \frac{1}{2} \ \pa_{\mu} \Phi^T \pa^{\mu} \Phi + \ov{\mu}^2 (\Phi^T \Phi)- \lambda (\Phi^T \Phi)^2 \ee
where $\ov{\mu}^2$ (not to be confused with the renormalization scale $\mu$, to be introduced in Eq.~(\ref{IM})) is positive (opposite in sign to a scalar field mass term) and $\lambda > 0$. With this choice of parameters the L$\sigma$M presents  spontaneous symmetry breaking from $SO(N+1)$ to $SO(N)$. The field $\Phi$ acquires a vacuum expectation value (VEV) where the field configuration of minimum energy verifies (at tree level)
\be \Phi^T \Phi = \frac{\ov{\mu}^2}{2 \lambda} = f_{\pi}^2 = NF^2.\ee
Denoting $\pi^a \ (a=1,...,N)$ to the $N$ first components of $\Phi$ and $\sigma$ the $N+1$ component we can choose
the VEV in the direction of the latter. Thus we have $\la \pi^a \ra=0$ but $\la \sigma (T=0) \ra = f_{\pi}$. The pions are the $N$ massless Nambu-Goldstone bosons; on the other hand, the field $\sigma$ acquires a mass equal to $m_{\sigma}^2=8 \lambda NF^2$.
Taking the limit $m_{\sigma}^2 \rightarrow \infty$ one can express $\sigma$ in terms of the pions as $\sigma=\sqrt{f_{\pi}^2-\pi^a \pi^a}$. This is the nonlinear sigma model in which one can eliminate explicitly the $\sigma$ degree of freedom.

Moreover, we can insert a possible $\pi$--mass term that explicitly breaks the $SO(N+1)$ symmetry in the direction of $\sigma$. This term should be in the form $\mathcal{L}_{\textrm{SB}}=c \sigma$, with $c$ a dimensionful constant. Expanding $\sigma$ in terms of the pions and requiring the appropriate form of the pion-mass term one can deduce that $c=m_{\pi}^2 f_{\pi}$. Explicitly,
\be \mathcal{L}_{\textrm{SB}} \equiv  m_{\pi}^2 f_{\pi} \sigma = m_{\pi}^2 f_{\pi}^2 \sqrt{1-\frac{\pi^a \pi^a}{f^2_{\pi}}} = m_{\pi}^2 f^2_{\pi} -\frac{1}{2} m_{\pi}^2 \pi^a \pi^a+\cdots  \ee

In the rest of the paper we will assume $F$ is a constant and denote the pion decay constant or the $\sigma$ VEV at $T=0$ as $f_{\pi}$ (it implicitly depends on $N$ because $f_{\pi}=\sqrt{N} F$). At finite temperature we will employ $v(T)=\la \sigma (T) \ra$ indistinctly. In particular, note that $v^2(0)=NF^2=f^2_{\pi}$.

The phase diagram can be understood by computing the effective potential $V_{\textrm{eff}}$ at finite temperature as a function of the $\sigma$-condensate, to ascertain whether and when the minimum of the potential corresponds to a symmetry--breaking phase. 

This effective potential is sometimes extracted from the generating functional of 1PI, $n$-point Green functions, $\Gamma (\phi,G^{-1})$ (see \cite{Cornwall:1974vz} for details). Here, we follow a different (but related) approach. The key formulas for understanding the procedure are given in Appendix~\ref{app:effVmath}.

Once computed, the effective potential is a function of the $\sigma$-field, an auxiliary inverse propagator $G^{-1}[0,\chi]$ and the temperature. It reads, in terms of renormalized quantities,

\be \label{effV}
 V_{\textrm{eff}} = \frac{1}{2} \left( \sigma^2 - NF^2 \right) G^{-1} [0,\chi] - \frac{(G^{-1}[0,\chi])^2}{16} \left( \frac{1}{\lambda_R} + \frac{N}{4\pi^2} \log \frac{\mu^2}{G^{-1}[0,\chi]}\right)+ \frac{N (G^{-1}[0,\chi])^2}{8(4\pi)^2}-m_{\pi}^2 f_{\pi} \sigma -  \frac{N}{2} g_0 (T, G^{-1}[0,\chi]).
\ee

Assuming a spatially homogeneous $\sigma$ condensate, the saddle-point approximation for this effective potential yields the value of $\la \sigma \ra$, together with a second equation of motion which is just the constraint necessary to solve for $G^{-1}[0,\chi]$, 
\be 
\label{saddle2} \frac{d V_{\tr{eff}}}{d \sigma
}=0; \quad \frac{d V_{\tr{eff}}}{d (G^{-1}[0,\chi] )}=0, 
\ee
where the analogy with the Cornwall-Jackiw-Tomboulis formalism is evident at this point (compare with Eqs.~(1.1a-1.1b) in \cite{Cornwall:1974vz}). Explicitly these equations read

\be \label{motion} \left\{ \begin{array}{l}
 0  =  \sigma \ G^{-1} [0,\chi] - m^2_{\pi} f_{\pi}, \\
 0  =  \frac{1}{2} \left( \sigma^2 -NF^2 \right) - \frac{G^{-1}[0,\chi]}{8} \left( \frac{1}{\lambda_R} - \frac{N}{4\pi^2} \log \frac{e \ G^{-1}[0,\chi]}{\mu^2} \right)+ \frac{N}{2} g_1(T,G^{-1}[0,\chi]).
\end{array} \right. \ee

The first one minimizes the potential in terms of $\sigma$. The second is the implicit constraint for $G^{-1}[0,\chi]$ for any value of  $V_{\textrm{eff}}$, not necessarily at the minimum. When $m_{\pi}=0$ the first equation in (\ref{motion}) reveals that there exist one phase with $\sigma=0$ and $G^{-1}[0,\chi] \neq 0$ and another phase where $G^{-1}[0,\chi]=0$ and $\sigma \neq 0$. In the latter case, the second equation yields
\be \sigma^2(T) = NF^2 - N g_1(T,0), \ee
and finally using the result in (\ref{g1chiral}) one arrives at
\be \label{condensate} \sigma(T) =\sigma_0 \left( 1- \frac{T^2}{T^2_c} \right)^{1/2}, \ee 
with $\sigma_0^2 \equiv f^2_{\pi}$ and the critical temperature 
\be \label{criticalT} T_c^2 \equiv 12 F^2. \ee
One can also check that the critical exponent for the order parameter is equal to the one from mean-field theory $\beta=1/2$.

The case with $m_{\pi} \neq 0$ is not so easily tractable and must be solved on a computer by an iterative method, such as Newton's method. Convergence is easily achieved.

\begin{figure}
\includegraphics[height=3in]{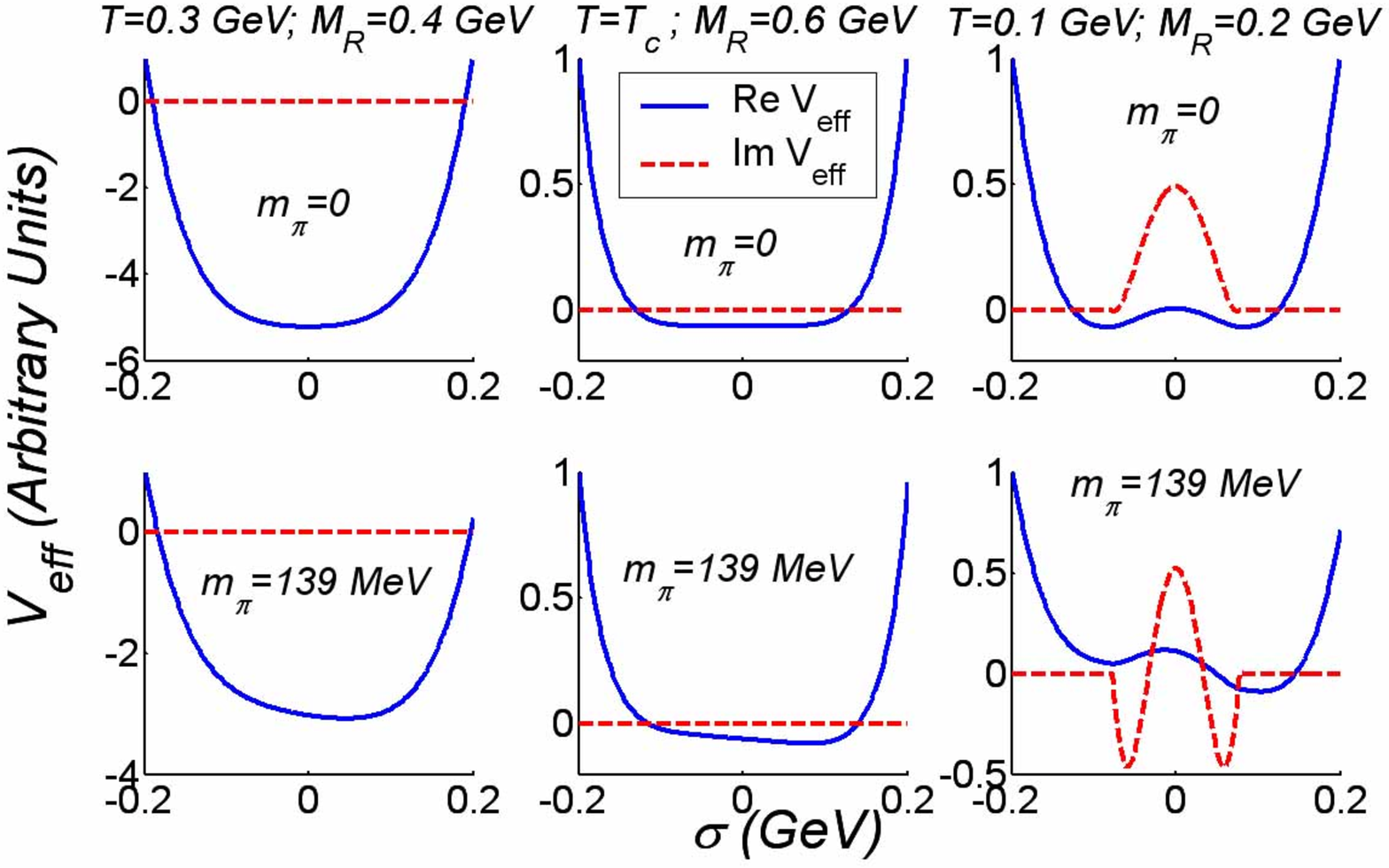}
\caption{Effective potential in various regimes of the L$\sigma$M, showing the various phases and the nature of the phase transitions between them. Top row: massless pions. Bottom row: pion mass fixed at $139$ MeV. From left to right, the renormalized $\sigma$ mass and temperature are varied as indicated in the graph. Each potential has been multiplied by an arbitrary temperature-independent constant factor, to match their scales for visibility. See the main text for interpretation of the results.
\label{fig:Vs}}  
\end{figure}

\begin{figure}
\includegraphics[height=2.2in]{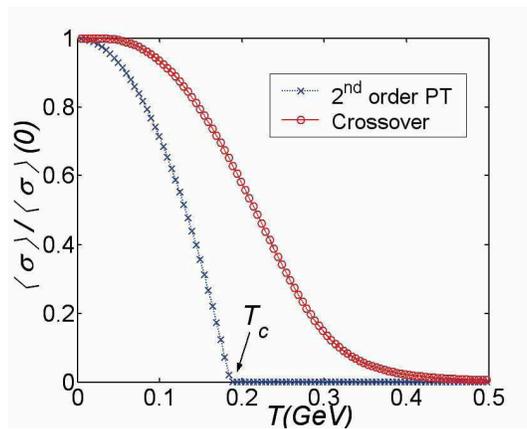}
\caption{The value of the $\sigma$ condensate as a function of temperature. Setting $m_\pi=0$, the model presents a clear second-order phase transition (PT) with a discontinuity in the derivative. For finite pion masses, we have a crossover.
\label{fig:crossover}}  
\end{figure}

Eliminating $G^{-1}[0,\chi]$ this way, one can easily plot the effective potential as a function of $\sigma$ at fixed temperature (Fig.~\ref{fig:Vs}). Identifying the position of its minimum with the help of a computer, we can also plot the value of the $\sigma$ condensate at the minimum as a function of $T$ (see Fig.~\ref{fig:crossover}).

In Fig.~\ref{fig:Vs}, we obtain the shape of the effective potential as a function of the relevant parameters, temperature $T$, renormalized mass of the $\sigma$, $M_R$ \footnote{$M_R$ can be traded for the renormalized coupling constant $\lambda_R$, through Eq.~(\ref{mrandlr})} and the physical mass of the pions $m_{\pi}$ (analogous to the coupling of an external field that explicitly breaks the original symmetry). For typical values of these parameters see Table~\ref{table1}.

\begin{table}
\caption{\label{table1} Default parameters used in the computation of $\eta/s$ in the L$\sigma$M (used unless noted otherwise).}
\begin{ruledtabular}
\begin{tabular}{cc}
Parameter & Value \\
\hline
$m_{\pi}$ & $139.57$ MeV\\
$f_{\pi}$ & $93$ MeV\\
$N$ & $3$ \\
$M_{R0}$ & $0.5$ GeV \\
$\lambda_R$ & $M_{R0}^2/(8f^2_{\pi}) \sim 3.6$ \\
$\mu^2$ & $1$ GeV$^2$\footnote{Nothing depends on this scale choice when using (\ref{runningmr}) and (\ref{mrandlr}).}\\
$T_c$ & $2\sqrt{3} F = 0.186$ GeV \\
\end{tabular}
\end{ruledtabular}
\end{table}

Inspection of the top row in Fig.~\ref{fig:Vs} reveals  a second-order phase transition from a potential with a single minimum at $\sigma=0$ at high temperature, to a phase with spontaneously broken symmetry below the critical temperature, where $\la \sigma \ra \neq 0$.

An imaginary potential $\textrm{Im} \ V_{\textrm{eff}} \neq 0$ arises when the real part $V_{\textrm{eff}}$ becomes convex, with negative second derivative. This is a well-understood phenomenon \cite{Weinberg:1987vp} and is related with the probability of decay per unit volume of the unstable vacuum.   

In the bottom row we consider the effect of a finite pion mass. Then the potential loses the residual left-right symmetry (reflection respect to  the $\sigma=0$ axis), and the pion mass is acting as an external magnetic field in an analogous condensed matter system.
The second-order phase transition becomes then a smooth crossover.

In order to further characterize the phase transition we now turn to Fig.~\ref{fig:crossover}, with the dependence of $\sigma$ taken at the minimum of $V_{\textrm{eff}}$, as a function of $T$. This is nothing but the order parameter of the system. The two possibilities in this model are now clear: a second-order phase transition and a crossover, if we introduce an explicit pion mass. Note that only the positive VEV is depicted in Fig.~\ref{fig:crossover}.

The phase transition in this model is a chiral one, where the pion acquires a thermal mass above the critical temperature. This continuously matches to zero at the critical temperature, and we cannot show a first order phase transition for any combination of the parameters \footnote{One can artificially force it by introducing an ad-hoc discontinuity in $m_\pi$ in the computer code, and we have examined this case to check that $\eta/s$ inherits the same discontinuity.}.

Combining the knowledge of $V_{\textrm{eff}} (\sigma)$ and $\la \sigma (T) \ra$ we are able to determine the critical temperature of the chiral transition and more generally, the entire phase diagram of the theory.

\section{Computation of the KSS number: $\eta/s$}\label{sec:KSS}

In this section we quote our computations of $\eta/s$ for the L$\sigma$M at large--$N$. To compute the shear viscosity we employ two different methods.

The first calculation  uses the quantum transport equation of traditional kinetic theory. Many details for this Boltzmann--formulation are given in our other publications \cite{Dobado:2003wr, Dobado:2007cv} and we omit them here for shortness. In  Appendix~\ref{kineticvisco} we sketch a few  steps for completeness.
The Boltzmann-Uehling-Uhlenbeck is a transport equation for the distribution function of the gas. This equation is linearized and solved perturbatively in an appropriate polynomial basis, in the so-called Chapman-Enskog formalism. The shear viscosity of the gas is then easy to obtain projecting the solution.

An alternative method to calculate the shear viscosity is the Green-Kubo equation.  This method, based on Linear Response Theory, works better for a field theory that accepts a diagrammatic expansion. Some details of the Green-Kubo formalism are reviewed in the Appendix~\ref{kubovisco}.
Consistency of both approaches provides a check of the calculation through kinetic theory and field theory methods. Results should be equivalent as has been shown in~\cite{Jeon:1995zm}.

A further check is provided in Appendix \ref{app:density}, where we revise the validity of the Boltzmann approximation to out of equilibrium computations. In essence, this ``molecular chaos'' hypothesis needs successive collisions for the same particle to be uncorrelated, requiring small densities as compared to the typical scattering cross section, $n \sigma^{3/2}<<1$. This we establish in that Appendix for the range of validity of our study.

The other ingredient needed for $\eta/s$ is the entropy density, that can also be calculated in two different ways. Frequently quoted is the free entropy of a degenerate Bose gas. For vanishing chemical potential one can obtain a simple formula for the entropy density from the Bose-Einstein partition function
\ba \log Z = -NV \int \frac{d\mathbf{p}}{(2\pi)^3} \log \left( 1-e^{-\beta E_p}\right), \\
s (T)\equiv \frac{\pa}{\pa T} \left( \frac{T}{V} \log Z \right)\ 
\ea
For massless pions, this simply reduces to 

\be 
\label{entropiatrivial}
s(T)=\frac{2N}{45}\pi^2 T^3\ .
\ee

An \emph{a priori} better and more accurate alternative is the entropy density derived from the effective potential in Eq.~(\ref{effV}). One should then regard $V_{\textrm{eff}}$ as a Helmhotz free energy obtained from the generating functional (as a partition function). 
\be s(T) = - \frac{\pa V_{\textrm{eff}}}{\pa T} = \frac{N}{2} \frac{\pa g_0 (T,G^{-1} [0, \chi])}{\pa T}.\ee

The limit of Eq.~(\ref{entropiatrivial}) for noninteracting massless pions is recovered upon employing Eq.~(\ref{g0chiral}).
We again resort to a computer for the, more complicated general case; our numerical evaluation however shows that the entropy remains close to the free gas, since at moderate temperatures the density is small. 

\begin{figure}
\includegraphics[height=2in]{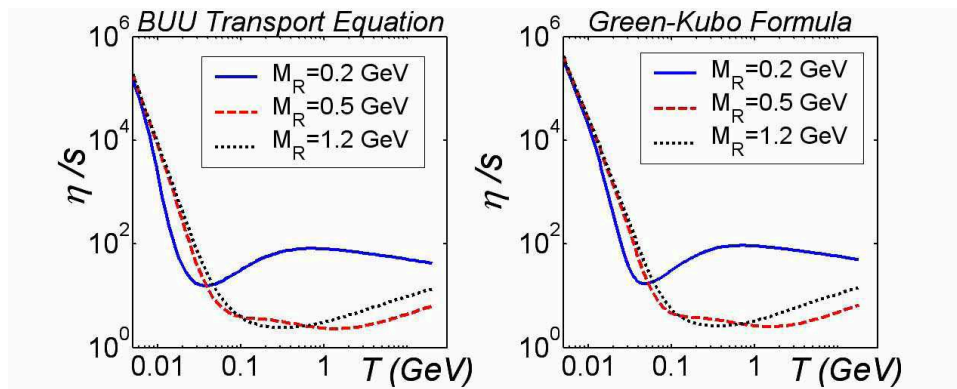}
\caption{Comparison of the viscosity computed with the Boltzmann-Uehling-Uhlenbeck equation (left panel) and the Green-Kubo formalism (right panel). The result is nicely consistent for a broad range of temperatures and renormalized masses.
\label{fig:comparametodos}}  
\end{figure}

Combining the results for $\eta$ and $s$ we can deduce $\eta/s$ as a function of temperature and plot it in Fig.~\ref{fig:comparametodos}. Both formalisms are shown (left panel from Boltzmann equation, right panel from Green-Kubo formula) and one can see the excellent agreement between them. 

The last step in our analysis is to connect the results of $\eta/s$ and those for the critical temperature of the L$\sigma$M.

\section{Minimum of $\eta/s$ and variation of the
$\la \sigma \ra$ condensate}\label{sec:minimum}

\begin{figure}
\includegraphics[height=2.4in]{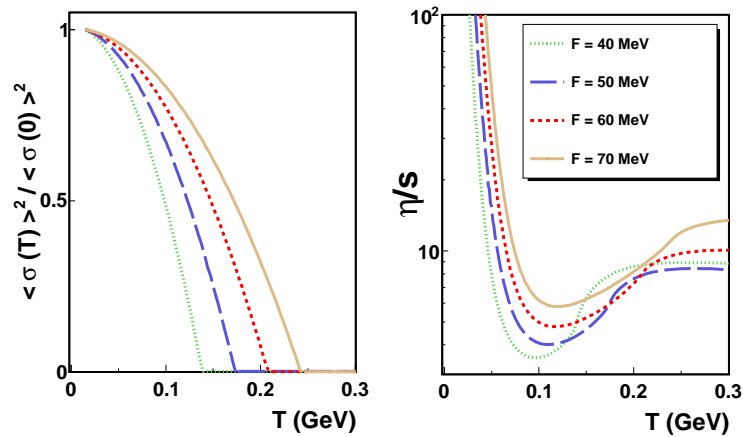}
\caption{We follow the evolution with $F$ of the minimum of $\eta/s$ and the $\sigma$ VEV, that yields the critical temperature at vanishing condensate. Both increase the same relative amount when varying the coupling constant. 
\label{fig:Fdependence}}  
\end{figure}

We now establish that the minimum of $\eta/s$ tracks the movement of the critical temperature when the model parameter $F$ is varied. For this we return to the case $m_\pi=0$ so as to have a second-order phase transition in the L$\sigma$M.
Recalling Eq.~(\ref{criticalT}) the critical temperature  is 
\be \label{Tcdependence} T_c = 2 \sqrt{3} F\ ,
\ee 
Therefore we need to extract the dependence of $\eta/s$ on $F$. This is done in the right panel of Fig.~\ref{fig:Fdependence}. For ease of comparison, we show in the left panel how $T_c$ depends on $F$. 
For the studied range of $F$ the critical temperature (defined as the temperature at which $\la \sigma \ra$ reaches a zero value), extracted from the computer code, varies exactly like the analytical Eq.~(\ref{Tcdependence}), which provides a handy check.

 First, we note that $\eta/s$ presents a nonanalyticity as a function of $T$ at $T=T_c$ where a very different qualitative behavior of $\eta/s$ begins. This is due to the dependence of $\eta/s$ on the  thermal pion mass $m_\pi(T)$ and the $\sigma$ condensate, both nonanalytical at $T_c$. 
Below $T_c$, the thermal pion mass  is identically zero where Goldstone's theorem protects the pion masslessness from radiative corrections. In this phase the symmetry is still broken and the condensate $\la \sigma \ra$ is nonzero and described by Equation (\ref{condensate}). Above $T_c$ the symmetry is restored and $\la \sigma \ra$ vanishes. The ``classical'' pion mass $m_{\pi}$ is zero as well. However, the quantum thermal corrections force a temperature-dependent thermal mass, $m_{\pi} (T)$ that is obtained from the thermal tadpole in the pion self-energy. Thus we have
\ba
\la \sigma(T)\ra = 
\left\{ \begin{array}{lr}
 \sigma(T) & T < T_c \\
0 & T \ge T_c
\end{array}
\right. 
\\
 m_{\pi}^2(T)= 
\left\{ \begin{array}{lr}
 0 & T < T_c \\
\frac{N}{3} \lambda_R (T^2-T^2_c) & T \ge T_c
\end{array}
\right. 
\ea
that is continuous at $T_c$ but nonanalytic. The behavior of $\la \sigma \ra$ itself is also continuous as shown in Fig.~\ref{fig:crossover} but with a discontinuous derivative at $T_c$. These two quantities influence both $\eta$ and $s$ and their nonanaliticity is inherited by the KSS ratio.

By varying $F$, the minimum of $\eta/s$  moves according to the $F$ dependence of $T_c$, providing evidence for our claim that $T_c$ and the minimum of $\eta/s$ are related.

\begin{figure}
\includegraphics[height=2.8in]{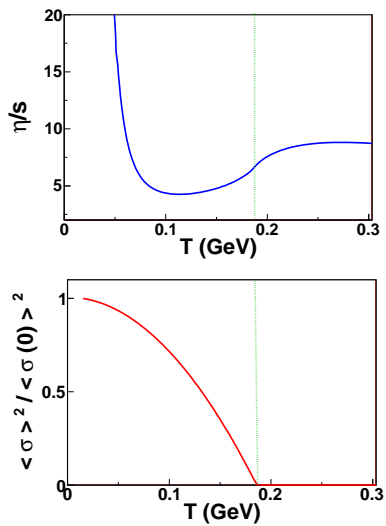}
\caption{Our key result: the minimum of $\eta/s$ occurs just before the critical temperature for the phase transition in the Linear Sigma Model. This is where the condensate varies rapidly approaching zero. The phase transition is marked by a nonanalyticity of the ratio.
\label{fig:minimumTc}}  
\end{figure}

In Fig.~\ref{fig:minimumTc} we zoom in the $\eta/s$ plot near the critical temperature.
As can be seen, the minimum is not reached at the critical temperature, but right before $T_c$. This shows that the minimum of $\eta/s$ is controlled  by the rapid variation of the order parameter. To understand this result one needs to keep in mind the diffusive nature of the transport in a gas. 
With increasing temperature, the gas particles carry transverse momentum between different parts of the gas more efficiently, and thus increase the shear viscosity. However their interactions hamper transport. This can be seen from Eq.~(\ref{pipiamp_F}). As $\la \sigma\ra$ decreases rapidly, the pion elastic cross section increases. Since $\eta\propto 1/\sigma_{\pi\pi}$ in kinetic theory, the viscosity must drop. Eventually growth is regained as the temperature increases.

Although the L$\sigma$M does not present a first order phase transition, within our treatment one can artificially force it by adding a jump for $\la \sigma \ra$ in the  computer code. In this case, the viscosity over entropy density ratio also turns out to have a discontinuity in the program (not shown since it is not a genuine model prediction) with a minimum of $\eta/s$ at exactly $T_c$ due to the jump, analogous to the first order transition in atomic Argon in our prior work \cite{Dobado:2008vt}.

\section{Critical exponent for $\eta/s$}\label{sec:critical}

We have argued \cite{Dobado:2008jr} that $\eta/s$ has a tiny critical exponent of about $y \simeq 0.04$ at the critical end-point of QCD by applying the theory of dynamic critical phenomena \cite{Hohenberg:1977ym}. This is true only if the dynamical universality class of the QCD critical point is that of Model H, as claimed in \cite{Son:2004iv}. The critical exponent $y$ (sometimes called $\phi$) is defined near the critical temperature as
\be \eta \propto |t|^{-y}, \ee 
with $t=(T-T_c)/T_c$ being the reduced temperature. 
The shear viscosity and the correlation length $\xi$ are related by a further power-law with exponent $x_{\eta}$  
\be \eta \propto \xi^{x_{\eta}}. \ee
Because $\xi \propto |t|^{-\nu}$ one can relate both dynamical exponents through the static one $y=x_{\eta} \nu$.

The critical exponent $x_{\eta}$ appears to be too small (near $0.06$) to be detected experimentally in a collision of heavy ions, a rather noisy environment. For this reason we expect that experiments will bear a residual minimum (as a function of temperature) in $\eta/s$, near the critical end--point. 

This does not mean that the critical behavior is unobservable. On the other hand, once measurements of the volume viscosity become available, its own critical exponent, near to 3 \cite{Moore:2008ws}, might well be observable.

Another not-so-small critical exponent to detect is $z$, defined as $\tau \propto \xi^z$, where $\tau$ is the relaxation time and $\xi$ the correlation length. Near the critical point, $\xi$ diverges and the previous formula reflects the so-called ``critical slowing-down''. 

It is interesting to note that the hyperscaling expression \footnote{We thank Victor Martin Mayor for the observation.} in \cite{Hohenberg:1977ym, Son:2004iv} relates the critical exponents for the shear viscosity and heat conductivity $x_{\eta}+x_{\lambda} = 4 -d - \eta$.
Further employing the relation $z=4-\eta-x_{\lambda}$ one obtains a relation between dynamical critical exponents:
\be \label{hyperscaling} x_{\eta} = z-d,\ee
where $d$ is the spatial dimension. For the QCD critical point, belonging to Model H one sees that $z \gtrsim 3$ and $x_{\eta}$ is  nearly zero. An estimate of $x_{\eta}$ might be obtained with the assistance of Eq.~(\ref{hyperscaling}), but needing a very precise measurement of $z$ through the long-time behavior of the relaxation process.

It is highly possible that a few modes alone are responsible for this theorized divergent behavior of the viscosity at the critical end-point, while other system modes yield simply a broad minimum. This separation is sometimes referred to \cite{Senders:1985}
as ``regular viscosity'' and ``singular viscosity''. In any case, from the L$\sigma$M side, mean-field theory (sort of what one recovers in large $N$) loses all information about the critical exponent (remember for example the 3D Ising model where the critical exponent for the heat capacity is $\alpha\simeq 0.11$ and mean-field theory yields zero).

In spite of the presumed critical behavior of shear viscosity in systems belonging to Model H (such as, presumably, QCD, but not the L$\sigma$M here treated), there exist many systems where $\eta$ does not diverge at the critical point. The divergence of $\eta/s$ is therefore not truly universal at the critical point, it depends on which universality class the system belongs to. The presence or absence of this and other critical exponents is important to help us classify the universality class and dynamical model of the QCD phase transition.
\subsection{The case of Helium-4}

\begin{figure}
\includegraphics[height=2.5in]{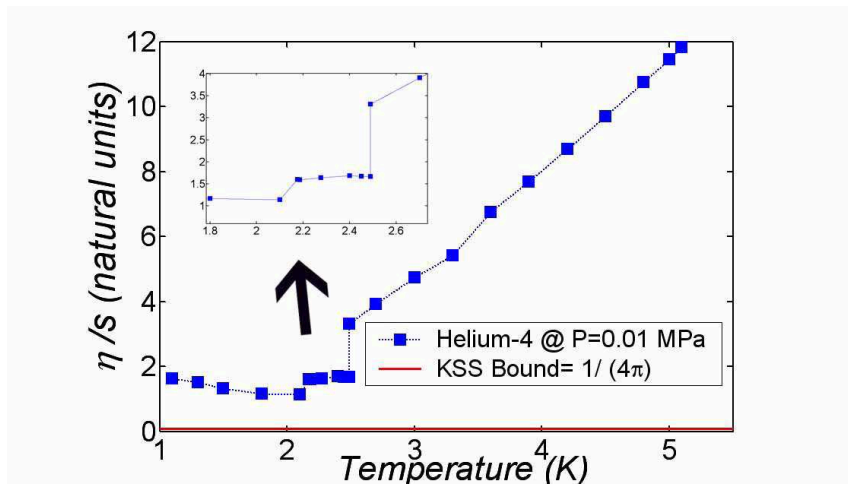}
\caption{$\eta/s$ for the three phases of helium-4 at low pressure, namely, He-II, He-I and
gaseous helium.
\label{fig:helium}}  
\end{figure}

As an example, we will bring forth the superfluid phase transition of Helium--4 at the $\lambda$-point. Its dynamic universality class is that of Model F, in which the order parameter is taken to be the macroscopic wave function, $\psi$. The quantity $n\equiv |\psi|^2$ tells about the concentration of quasiparticles belonging to the condensate.  In this universality class \cite{Hohenberg:1977ym} $\eta$ is not singular at $T_c$. 
Without passing by the critical end-point, the behavior of $\eta/s$ as a function of temperature is shown in Fig.~\ref{fig:helium}. The data are experimental measurements for gaseous Helium, normal liquid He-I and superfluid He-II. We first point out that the KSS bound is always preserved, even at $T_{\lambda}$. Below $T_{\lambda}$ the superfluid is formed by the condensation of Bogoliubov quasi--particles. The nonzero value of $\eta$ and $s$ in He-II can be understood with Landau's ``two-fluid'' model. Both coefficients approach zero when $T$ descends towards absolute zero, but the behavior of $\eta/s$ at low temperatures is to increase as observed early--on \cite{Kovtun:2004de}.

More interesting is that two apparent local minima occur at both critical temperatures (superfluid $T_{\lambda}$ and vaporization $T_c$). A double minimum would entail that $\eta/s$ could have a maximum between $T_{\lambda}=2.177$ K and $T_c=2.489$ K, in the He-I phase. (This phase is not a conventional liquid, the viscosity itself is anomalous and it softly decreases when temperature decreases). Due to the low resolution of the experimental data, clear minima are not seen, it is interesting to conjecture whether this behavior could be truly universal, i.e., independent of the dynamical universality class, with more generality than our theoretical considerations based on the $\sigma$ model. This is left for future work.

\section{Summary and outlook}\label{sec:conclusions}

In this paper we have been able to obtain the entire phase diagram as function of $(T,M_R,m_\pi)$ of the L$\sigma$M in the large--$N$ limit by means of the effective potential of the theory. We have located the critical temperature where the chiral phase transition takes place. The temperature dependence of the order parameter, i.e., the condensate $\la \sigma \ra$ has also been showed.
 
We have calculated the KSS number, or $\eta/s$, for this model as a function of temperature using both quantum kinetic theory and the Green-Kubo formalism. A very good agreement between them has been obtained for every value of $M_R$, $m_{\pi}$ and other parameters of the model. 

Combining both results we have shown that a strong dependence indeed exists  between the minimum of $\eta/s$ and the phase transition of the L$\sigma$M at large--$N$. This minimum is not achieved exactly at $T_c$ where the phase transition occurs, but a little before, due to the falling of $\la \sigma \ra$ towards zero. 

We also would like to discuss the generality and reach of the result. The minimum of $\eta/s$ near the phase transition does not depend on any parameter exclusive of the L$\sigma$M. It just depends on the falling behavior of $\la \sigma \ra$ from a nonzero value at $T=0$ to zero at $T_c$. The VEV of the $\sigma$ field can be understood as the order parameter of the chiral phase transition, and every other phase transitions has an equivalent order parameter (spontaneus magnetization in spin systems, macroscopic wave function in superfluids and superconductors...). 
If this order parameter influences the transport coefficient, one should expect our conclusion to be generalized in a straightforward way.

\begin{figure}
\includegraphics[height=2.3in]{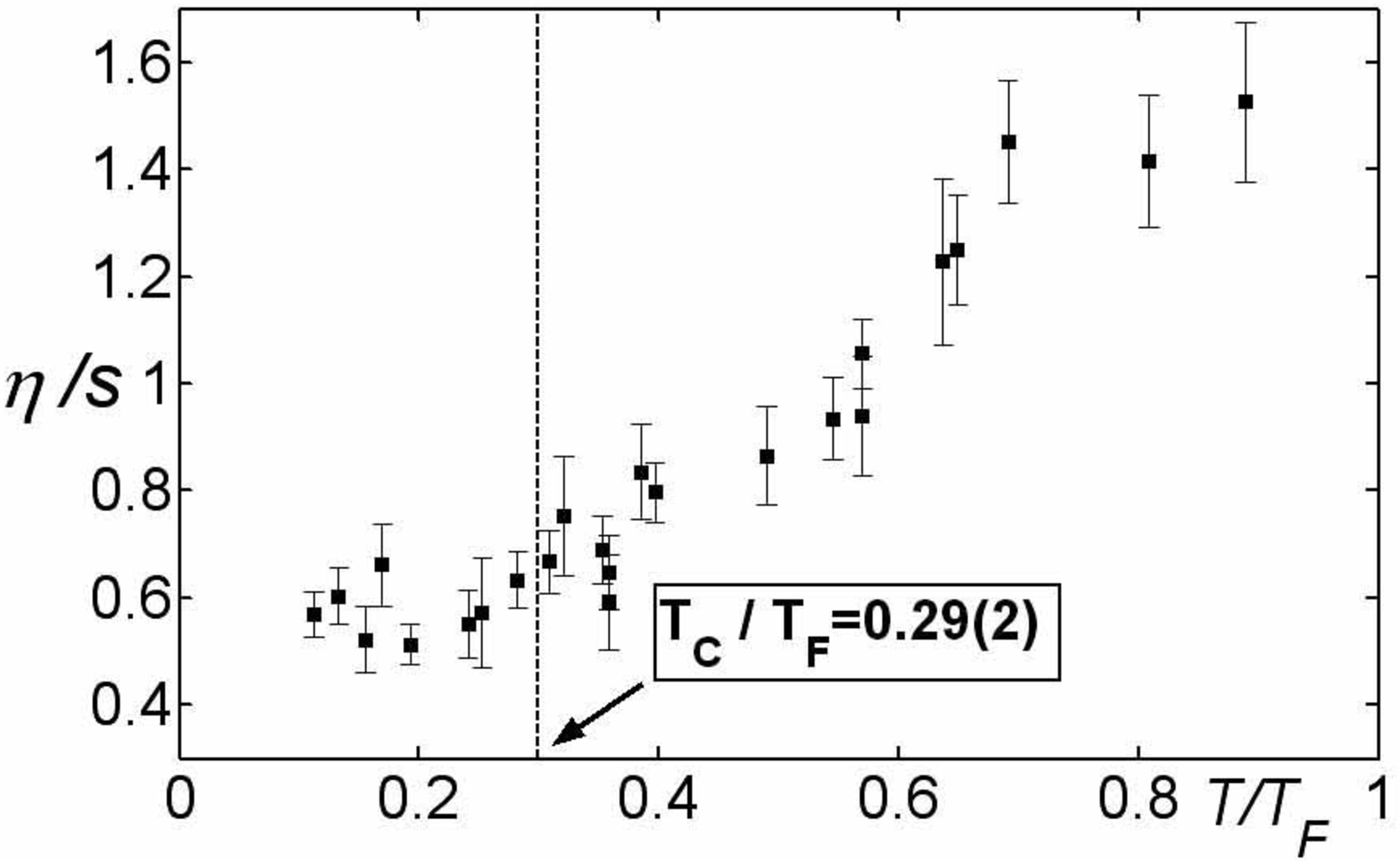}
\caption{The plot from \cite{schaefer-2007-76} shows that the minimum of $\eta/s$ for a trapped cold atomic gas in the unitary limit appears slightly before the phase transition temperature $T_c=0.29(2)T_F$. A simple quadratic fit of the data shows a minimum just below $T=0.2T_F$ with a $\chi^2/d.o.f=0.071$ whereas if we constrain the minimum at $T_c/T_F$, the quadratic fit has an almost six times larger $\chi^2/d.o.f=0.39$.}
\label{Coldfermiatoms}
\end{figure}

We have scanned the recent literature to see if other fluid systems
that have been considered in connection with the KSS conjecture have a minimum value of $\eta/s$ that occurs at a temperature discernibly smaller than the phase transition. This behavior is indeed present already in \cite{schaefer-2007-76}, but it seems  to have passed unnoticed. In Fig.~\ref{Coldfermiatoms} we replot the data given in \cite{schaefer-2007-76} but add the location of the critical temperature claimed by the same experimental group that provided the original data. The minimum of $\eta/s$ for this cold Fermi gas near the unitary limit occurs slightly before the critical temperature. We look forward to other examples where enough precision can be achieved to separate the phase transition and the minimum of the viscosity.

Meanwhile, one could ask what the situation for Yang-Mills theory is (as relevant to the Quark-Gluon-Plasma). The perturbative computations by other authors that we have quoted~\cite{Arnold:2003zc} are not of much use near $T_c$. Using a nonperturbative computation, Antonov has shown~\cite{Antonov:2009xn} that a possible minimum should occur in the range $2T_c-3T_c$, definitely above the phase transition. This could fit our observation that a significant change of the order parameter can mark the minimum of $\eta/s$, as it has been pointed out by 
Hidaka and Pisarski~\cite{Hidaka:2008dr} that the Polyakov loop acts as an order parameter and influences $\eta/s$. The Polyakov loop only reaches a value near 1 at about $2T_c$.

\acknowledgments
The authors thank A. G\'omez Nicola, D. Fern\'andez Fraile, F. Sols and D. Antonov for useful discussions as well as T. Schaefer and N. Demir for providing us their $\eta/s$ data. This work was supported by grants No. FPA2007-29115-E, No. PR34-1856-BSCH, No. FPA2008-00592, No. FIS2008-01323, No. UCM-BSCH GR58/08 910309, UCM-Santander No. PR34/07-15875, and an FPU grant to J. M. T-R.

\appendix
\section{The effective potential for the linear sigma model at large--$N$}
\label{app:effVmath}

Taking the original Lagrangian in (\ref{lagrangian}) and performing a Wick rotation ($t \rightarrow -i \tau$), the Euclidean or imaginary time Lagrangian is (in the following we denote $\Phi^2\equiv \Phi^T \Phi$)
\be -\mc{L}_E [\Phi,\pa_{\mu} \Phi]=\frac{1}{2} \pa_{\mu} \Phi^T \pa^{\mu} \Phi-\ov{\mu}^2\Phi^2 + \lambda (\Phi^2)^2 - \mathcal{L}_{\textrm{SB}},\ee
where now the product in the $\mu$-index is to be understood as Euclidean.
The partition function reads
\be Z=\int [d\Phi] \ \exp{\left(-S_E[\Phi]\right)}=\int [d\pi^a] [d\sigma] \exp{\left( - S_E [\pi^a,\sigma]\right)}.
\ee

\begin{figure}
\includegraphics[height=1in]{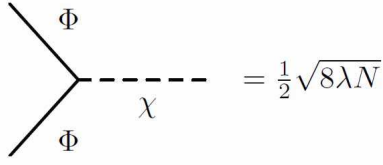}
\caption{Trilinear coupling between $\Phi$ and the auxiliary field $\chi$ that replaces the original quartic interaction $\Phi^4$.}
\label{fig:newcoupling}
\end{figure}

We wish to obtain an effective potential for the $\sigma$ field that allows us to decide under what circumstances one is in a condensed phase. For this we need to perform the integration over the dynamical $\pi^a$ variables.
Since the action is quartic in them, we introduce an auxiliary static field $\chi$, with no kinetic term, that splits the quartic pion vertex.
To assist large--$N$ counting, it is convenient to include a factor of $N^{-1/2}$ in its definition,  
\be 
\chi \equiv \Phi^2 \sqrt{\frac{2\lambda}{N}} \ .
\ee
The quartic potential is then hidden in the identity
\bw
\be \exp{\left( -\lambda \int d^4x \ (\Phi^2)^2\right)}=\int [d\chi] \exp{\left[\frac{1}{2} \int d^4x \left( N\chi^2- \chi \Phi^2 \sqrt{8\lambda N}\right)\right]}
\ee 
\ew
that features a trilinear coupling instead and is therefore quadratic in the pion fields. This coupling is shown in Fig.~\ref{fig:newcoupling}. This method has also been used in the L$\sigma$M at $T=0$ in \cite{Coleman:1974jh}.

Then, the complete partition function becomes
\be Z=\int [d\pi^a] [d\sigma][d\chi] \exp \left( -S_{\tr{eff}} [\pi^a,\sigma, \chi ]\right)\ee
with the full (still Euclidean) effective action
\bw
\be
S_{\tr{eff}} [\pi^a,\sigma,\chi] = \int d^4 x \left[\frac{1}{2} (\pa_{\mu} \pi^a)^2+\frac{1}{2} (\pa_{\mu} \sigma )^2 - \ov{\mu}^2\pi^2-\ov{\mu}^2 \sigma^2 - \frac{1}{2} N \chi^2 + \frac{1}{2} \chi \pi^2 \sqrt{8\lambda N} + \frac{1}{2} \chi \sigma^2 \sqrt{8\lambda N}  - m_{\pi}^2 f_{\pi} \sigma \right]. 
\ee
\ew
Integrating by parts the $\pi^a$ kinetic term we obtain
\bw
\be
\nonumber S_{\tr{eff}}  [\pi^a,\sigma,\chi] =  \int d^4 x \left[\frac{1}{2} \pi^a (-\square_E- 2 \ov{\mu}^2 + \chi \sqrt{8\lambda N} ) \pi^a +\frac{1}{2} (\pa_{\mu} \sigma )^2 -\ov{\mu}^2 \sigma^2
 - \frac{1}{2} N \chi^2 
 + \frac{1}{2} \chi \sigma^2 \sqrt{8\lambda N}  - m_{\pi}^2 f_{\pi} \sigma \right].
\ee
\ew
To make contact with the Cornwall-Jackiw-Tomboulis formalism in \cite{Cornwall:1974vz}, it is useful to further trade $\chi$ by the amputated two-point function for the $\sigma$, that can then be used as the auxiliary variable,
\be G^{-1} [q,\chi] \equiv q^2-2\ov{\mu}^2 + \chi \sqrt{8 \lambda N}. \ee
We may then perform the path integration over the $\pi^a$ fields to obtain the effective action for the $\sigma$ alone,
\ba \int [d\pi^a] \exp\left[ -\frac{1}{2}\int d^4 x \ \pi^a (-\square_E + G^{-1} [0,\chi])\pi^a \right] \propto \int d^4x \ \left[ \det\left(  -\square_E+G^{-1}[0,\chi] \right)\right]^{-N/2} \nonumber \\
= \int d^4x \ \exp \left[ -\frac{N}{2} \textrm{ tr } \log \left( -\square_E + G^{-1} [0,\chi] \right) \right] =\int d^4 x \ \exp \left[ - \frac{N}{2} \int \frac{d^4q}{(2\pi)^3} \log G^{-1}[q,\chi]\right] \ .
\ea
Therefore the effective action to be employed in the generating functional
$ Z=\int [d\sigma] [d\chi] \exp{\left(-\Gamma_{\tr{eff}} [\sigma,\chi] \right)} $
is, introducing the finite--temperature Imaginary Time Formalism with Matsubara frequencies $\omega_{n} =\frac{2\pi n}{\beta}$ ($\beta=1/T$) ,
$\int d^4 q\to T\sum_n \int d^3q$:
\bw
\be \label{gammaeff}
\Gamma_{\tr{eff}} [\sigma,\chi]=\int d^Dx \left\{\frac{1}{2} (\pa_{\mu} \sigma)^2 - \frac{N}{2} \chi^2 + \frac{1}{2} \sigma^2 G^{-1}[0,\chi]- m_{\pi}^2f_{\pi}\sigma + \frac{N}{2} T \sum_{n=-\infty}^{\infty} \int \frac{d^3q}{(2\pi)^3} \log G^{-1}[q,\chi] \right\} .
\ee
\ew

From Eq.~(\ref{gammaeff}) we read--off the effective potential. Rewriting the $-N\chi^2/2$ term in terms of $G^{-1}[0,\chi]$
\bw \be \label{needed} \frac{N \chi^2}{2}=\frac{\left(G^{-1} [0,\chi] + 2\ov{\mu}^2\right)^2}{16 \lambda}=\frac{(G^{-1} [0,\chi])^2}{16\lambda}+ \lambda N^2F^4+\frac{NF^2 G^{-1} [0,\chi]}{2}\ ,\ee \ew
we obtain $V_{\rm eff}(\sigma,\chi)$,
\bw
\be  \label{effpotential} V_{\textrm{eff}} (\sigma,G^{-1} [q,\chi])= \frac{1}{2} \left( \sigma^2 - NF^2 \right) G^{-1} [0,\chi] - \frac{1}{16 \lambda} (G^{-1}[0,\chi])^2 - \lambda N^2F^4 -m_{\pi}^2 f_{\pi} \sigma +  \frac{N}{2} T \sum_{n=-\infty}^{\infty} \int \frac{d^3q}{(2\pi)^3} \log G^{-1}[q,\chi],
\ee
\ew

As is well known, the thermal loop integrals such as in Eq.~(\ref{effpotential}) contain no new divergences aside from those in the vacuum integral (the term $n=0$). In this case

\be \int_{\beta} d\tilde{q} \ \log \ (q^2 + M^2) \equiv T \sum_{n \in \mathbb{Z}} \int d^3 \mathbf{q} \ \log \ (\omega_n^2+\mathbf{q}^2+M^2)  = \frac{M^2}{2} I_{M} - \frac{M^4}{4(4\pi)^2} - g_0 (T; M^2),\ee
where the zero-frequency term in the Matsubara sum, the vacuum loop integral, carries the divergence, that one can extract in dimensional regularization with 
$ N_{\epsilon}=\frac{2}{\epsilon} + \log 4\pi -\gamma $,
\be \label{IM} 
I_{M}= -\frac{M^2}{(4\pi)^2} \left(N_{\epsilon} +1+ \log \frac{\mu^2}{M^2} \right)\ee 
and $g_0$ is the thermal part of the loop integral. These $g$--functions are those previously defined in \cite{Gasser:1986vb}. Another useful integral (needed in the saddle-point equations) is the following:
\be \int_{\beta} d\tilde{q} \ \frac{1}{q^2+M^2} =T \sum_{n \in \mathbb{Z}} \int d^3 \mathbf{q} \ \frac{1}{\omega^2_n + \mathbf{q}^2 + M^2}= I_{M} + g_1(T;M^2).\ee
We need one more thermal $g$--integral since Eq.~(\ref{motion}) is solved 
by iteration using Newton's method, which requires a derivative respect to $G^{-1}[0,\chi]$.
The series of $g$--integrals satisfy the recursion relation 
\be 
g_{n+1} = - \frac{dg_n}{dM^2}\ . 
\ee
In terms of the variable $y=M/T$, the ones we need are
\ba
 \label{g0} g_0 (T;M^2)&=& \frac{T^4}{3\pi^2} \int_y^{\infty} dx (x^2-y^2)^{3/2} \frac{1}{e^x-1},
\\ 
\label{g1} g_1 (T;M^2)&=& \frac{T^2}{2\pi^2} \int_y^{\infty} dx \ \frac{\sqrt{x^2-y^2}}{e^x-1},
\\
\label{g2} g_2 (T;M^2)&=&\frac{1}{4\pi^2} \int_y^{\infty} \ dx \ \frac{1}{\sqrt{x^2-y^2}} \frac{1}{e^x-1} \ .
\ea
In the limit $M\ll T$, or  $y \rightarrow 0$, one recovers the standard results for a massless Bose gas
\ba 
\label{g0chiral} g_0 (T,0) = \frac{T^4}{3\pi^2} \int_0^{\infty} dx \ \frac{x^3}{e^x-1} =\frac{T^4}{3\pi^2} \Gamma(4) \zeta(4) = \frac{T^4}{3\pi^2}6\frac{\pi^4}{90}=
\frac{\pi^2 T^4}{45}
\\
\label{g1chiral} g_1 (T,0)= \frac{T^2}{2\pi^2} \int_0^{\infty} \frac{dx \ x }{e^x-1} =\frac{T^2}{2\pi^2} \Gamma(2) \zeta(2) = \frac{T^2}{2\pi^2}\frac{\pi^2}{6} = \frac{T^2}{12} 
\ea
As the Bose-Einstein factors cut--off high momenta, only the vacuum parts contain  divergences, and these can be 
reabsorbed in the $\sigma$-mass and the coupling constant that appear in the classical Lagrangian density. In the next Appendix we describe the renormalization procedure.

\section{$\pi-\pi$ amplitude in the L$\sigma$M at large--$N$}
\label{pipiappendix}
In a nutshell, transport in a gas is a diffusive effect, so that in kinetic theory transport coefficients are proportional to the mean free path, or inversely proportional to the total collision cross--section. Thus we need the scattering amplitude in the L$\sigma$M in the large--$N$ limit.

\begin{center}
\begin{figure}[h]
\includegraphics[height=1in]{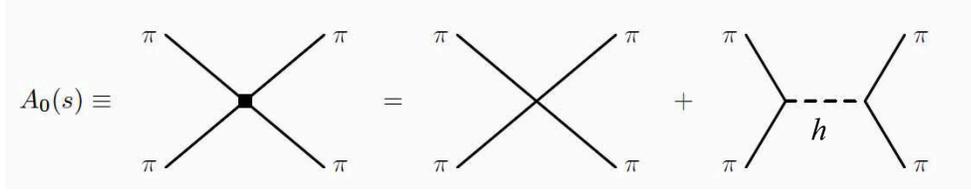}
\caption{Tree level $\pi-\pi$ amplitude at $\mathcal{O} (1/N).$}
\label{fig:amplitude0}
\end{figure}
\end{center}

This $\pi-\pi$ scattering amplitude at tree-level is simply the one shown in Fig.~\ref{fig:amplitude0}. 

\begin{center}
\begin{figure}[h]
\includegraphics[height=1in]{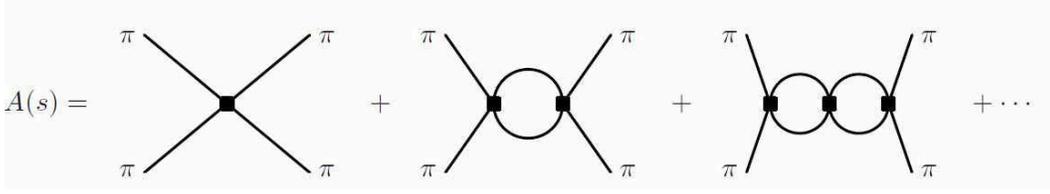}
\caption{Resummed amplitude for $\pi-\pi$ scattering at $\mathcal{O} (1/N).$}
\label{fig:amplitude}
\end{figure}
\end{center}

Combining both diagrams,
\be 
\label{diagramA0} A_0(s)= \frac{s}{v^2} \frac{1}{1-\frac{s}{M^2}},
\ee
where $v$ is  the VEV of the $\sigma$ field, that depends on the temperature $v(T)=\la \sigma (T) \ra$ and $M^2=m^2_{\sigma}=8\lambda v^2$ is the spontaneously generated mass of the longitudinal mode $h$. This exchanged boson $h$ is the physical quantum over the vacuum, $\sigma=v+h$.
In (\ref{diagramA0}) we explicitly see that $A_0$ is $\mathcal{O} (1/N)$. In the large--$N$ limit (with $NF^2$ fixed) the $s$-channel iteration of the tree level diagram is also of order $1/N$ (recall that every pion loop carries a factor $N$), and must be resummed. The situation is depicted in Fig.~\ref{fig:amplitude}.
The one-loop integral is $I(s)=\frac{1}{16\pi^2} \left[ N_{\epsilon} + 2 + \textrm{Log }(-\mu^2 / s) \right] $, where the Log is to be understood as complex and the integration has been regulated with the procedure of dimensional regularization. In this scheme one introduces a scale $\mu$ in the definition of the renormalized mass, $M_R(\mu)$ in order to absorb the infinity in $I(s)$.
 
Defining $M_R$ as the value of the renormalized mass at some predetermined energy $\mu_0$
\be 
M^2_R \equiv M^2_R(\mu_0),\ee
we can express the mass at an arbitrary renormalization scale $\mu$ 
\be \label{runningmr} M_R^2(\mu) = \frac{M^2_R}{1- \frac{M^2_R}{32 \pi^2 F^2} \log \left( \frac{\mu^2}{\mu_0^2} \right)}.
\ee
Not only the $\sigma$-mass is renormalized but also the coupling constant, $\lambda$. The dressed coupling constant is defined as
\be \label{lambdaR} \frac{1}{\lambda_R(\mu)} \equiv \frac{1}{\lambda} + \frac{N}{4\pi^2} \left(2+\log 4 \pi -\gamma +\frac{2}{\epsilon} \right).\ee

Both renormalized parameters are related (like their bare counterparts) through 
\be \label{mrandlr}
M^2_R(\mu)=8 \lambda_R(\mu) v^2.
\ee

The resummed amplitude in Fig.~\ref{fig:amplitude} becomes
\be 
A(s,t,u)=A(s) = \frac{A_0(s)}{1- \frac{NI(s)}{2} A_0(s)}=\frac{s}{v^2} \frac{1}{1-\frac{s}{M^2_R(s)}+\frac{sN}{32\pi^2v^2} \textrm{ Log } \left(\frac{-s}{\mu^2}\right)}\ \ .
\ee

The Mandelstam variable is positive $s >0$, so that choosing the branch cut of the logarithm along the positive $s$ axis, in order that $\textrm{ Log } (-s)=\log s + i \pi$, we arrive at
\be \label{pipiamp_F}
A(s) =\frac{s}{v^2} \frac{1-\frac{s}{M^2_R(s)} + \frac{sN}{32\pi^2 v^2} \log \left( \frac{s}{\mu^2} \right)-i \frac{sN}{32 \pi v^2}  }{\left[ 1-\frac{s}{M^2_R(s)}+\frac{sN}{32\pi^2v^2} \log \left(\frac{s}{\mu^2}\right) \right]^2+ \left( \frac{sN}{32 \pi v^2} \right)^2}\ \ .
\ee

The squared scattering amplitude is simply
 
\be \label{sqscamp}
|A(s)|^2 = \frac{s^2}{v^4} G(s),
\ee
in terms of the loop function
\be \label{loopfunction}
G(s)= \frac{1}{\left[ 1-\frac{s}{M^2_R(s)}+\frac{sN}{32\pi^2v^2} \log \left(\frac{s}{\mu^2}\right) \right]^2+ \left( \frac{sN}{32 \pi v^2} \right)^2}.
\ee

If we consider the partial amplitudes projected on the isospin channels we see that only $T_0$ dominates at order $1/N$:
\ba 
T_0 & = & NA(s)+A(t)+A(u) = N A(s) + \mathcal{O} \left( \frac{1}{N} \right) \\
T_1 & = & A(t)-A(u) = \mathcal{O} \left( \frac{1}{N} \right) \\
T_2 & = & A(t)+A(u) = \mathcal{O} \left( \frac{1}{N} \right)
\ea
It is also useful to construct the isospin-spin partial waves. They are defined as
\be t_{IJ}(s)=\frac{1}{64 \pi} \int_{-1}^{1} d(\cos \theta) T_I (s, \cos \theta) P_J (\cos \theta), \ee
where $P_J$ are the Legendre polynomials. 
For moderate energies ($|\mathbf{p}|<1$ GeV) we only keep the lowest partial wave in each isospin channel, namely $IJ=00,11,20$. In the limit $N\rightarrow \infty$ their $N$-dependence is
\ba
\label{tij} t_{00} (s) & = &\frac{N A(s)}{32 \pi}+ \mathcal{O}\left(\frac{1}{N} \right) , \\
t_{11} (s) & = & \mathcal{O} \left( \frac{1}{N} \right), \\
t_{20} (s) &= & \mathcal{O} \left( \frac{1}{N} \right),
\ea
so  $t_{00} (s)$ dominates the amplitude (note in true two-flavor QCD at $N=3$, $t_{11}(s)$ is also very strong due to the $\rho$ resonance).

For the case in which the pion mass is to be considered the amplitude $A(s)$ must be modified. The extended discussion of this case can be found in \cite{Dobado:1994fd}. The partial wave is expressed as
\be t_{00}(s)=\frac{N}{32 \pi} \left[ A_0(s)+A_m(s) \right],\ee
where the amplitudes are respectively
\ba 
\label{A0} A_0(s)=\frac{1}{v^2} \frac{s}{1-\frac{sN}{M^2_R(\mu)}-\frac{s}{32 \pi^2 v^2}T(s;\mu)},\\ 
\nonumber
\label{Am}  A_m(s)=-\frac{m_{\pi}^2}{v^2} \frac{1+\frac{2s}{M^2_R(\mu)}- \frac{sN}{16\pi^2 v^2} \log \left( \frac{m_{\pi}^2}{\mu^2} \right)}{\left[ 1-\frac{s}{M^2_R(\mu)} - \frac{sN}{32 \pi^2 v^2} T(s;\mu)\right]^2},
\\
T(s;\mu)\equiv \sqrt{1-\frac{4m_{\pi}^2}{s}} \log \left| \frac{\sqrt{1-\frac{4m_{\pi}^2}{s}}-1}{\sqrt{1-\frac{4m_{\pi}^2}{s}}+1} \right| + i \pi \sqrt{1-\frac{4m_{\pi}^2}{s}} - \log \left( \frac{m_{\pi}^2}{\mu^2} \right).
\ea

When $m_{\pi}=0$, the amplitude $A_m(s)$ vanishes and $T(s;\mu) \rightarrow \log  (-\mu^2 /s)$ and the $(IJ)=(00)$ partial amplitude reduces to
\bw 
\be \label{shift00}
|t_{00}|^2(s)=\frac{s^2N^2}{(32\pi v^2)^2} \frac{1}{\left[ 1- \frac{s}{M_R^2} + \frac{sN}{32\pi^2 v^2} \log \left(\frac{s}{\mu^2} \right)\right]^2+\left(\frac{sN}{32 \pi v^2}\right)^2}=\frac{s^2 N^2}{(32\pi v^2)^2} G(s),\ee \ew
with $G(s)$ defined in (\ref{loopfunction}). The same result could be obtained from $|t_{00}|^2$ in (\ref{tij}).

From Eq.~(\ref{tij}), the large--$N$ scattering amplitude $|\overline{T}|^2$ is given by the $s$-channel, $|\overline{T}|^2 \simeq |A(s)|^2$. The averaged cross section in the CM frame follows
\be \frac{d \overline{\sigma}}{d \Omega} = \frac{1}{2} \frac{|\overline{T}|^2}{64\pi^2 s}, \ee
or
\be\label{crosssection} \overline{\sigma}(s)=\frac{s}{32 \pi v^4} G(s)\ee
is the total cross section averaged over the final states.

On the other hand, the total cross section can be defined in terms of the isospin-spin projected scattering amplitudes as
\be \label{presigmatot}
\sigma_{\textrm{tot}}=\frac{32\pi}{3s} \sum_I\sum_J (2J+1)(2I+1) |t_{IJ}(s)|^2,
\ee

and the integrated total cross section, when only $(IJ)=(00)$ is considered, is
\be \label{sigmatot} \sigma_{\textrm{tot}} (s) \simeq \frac{32\pi}{3s} \frac{s^2 N^2}{(32\pi v^2)^2} G(s)=\frac{s N^2}{96 \pi v^4} G(s) \ee

Note that the Kubo formalism calls for the total cross section summed over isospins, while in Boltzmann's collision term the initial flavors are averaged, so there is a flavor factor between them
\be \frac{\sigma_{\textrm{tot}}}{\overline{\sigma}} = \frac{N^2}{3}. \ee

\begin{figure}
\includegraphics[height=2.8in]{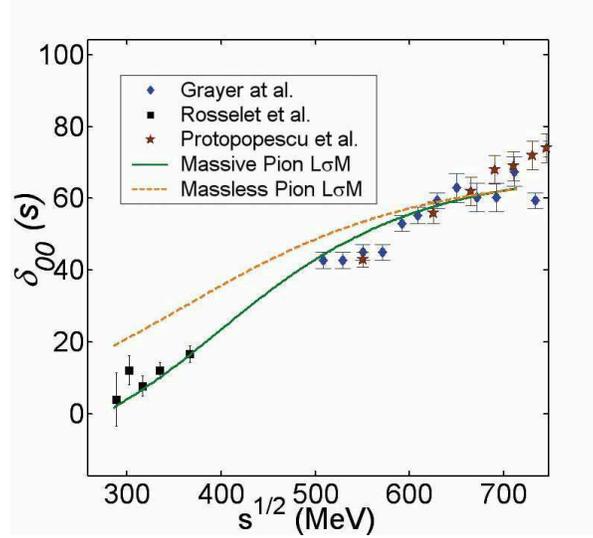}
\caption{\label{fig:phase} The scalar phase shift in the large-mass, large-$N$ limit of the  L$\sigma$M compared with the experimental data of \cite{Grayer:1974cr,Rosselet:1976pu,Protopopescu:1973sh}. The massless-pion curve (dashed  red on-line) corresponds to the amplitude given in Eq.~(\ref{A0}) and the massive-pion curve is obtained when adding Eq.~(\ref{A0}) and Eq.~(\ref{Am}). } 
\end{figure}

To compare with experimental $\pi-\pi$ data at $N=3$, the  scalar phase--shift $\delta_{00}$ corresponding to Eq.~(\ref{shift00}) is plotted in Fig.~\ref{fig:phase}, and reasonable agreement is seen. As discussed in \cite{Dobado:1994fd} data are well reproduced by taking $M_R\to \infty$, the only remaining parameters being $F$ and $\mu$. This renormalization scale is conventionally taken to be the $\rho$ mass, $\mu \simeq 775$ MeV, where one naturally expects (because we have neglected the contribution of the $(IJ)=(11)$ channel) that the L$\sigma$M ceases to be valid for $N=3$. 

An interesting point is to examine the breaking of scale invariance even at the critical temperature. This can be followed from Eq.~(\ref{shift00}) and (\ref{presigmatot}). All one needs to do is note that $\lim_{T\to T_c} v(T) = 0$ as the condensate vanishes, remembering also the relation of $M_R$ and $v$ through
Eq.~(\ref{mrandlr}). Then the cross section fails to scale as $\sigma \propto 1/s$ (conformality) due to the renormalization scale that brings in the usual logarithm of
$s/\mu^2$. This is simply a consequence of scale-invariance being anomalous. 

\section{Computation of the shear viscosity in kinetic theory}
\label{kineticvisco}

One approach to calculate the transport coefficients is based on kinetic theory by solving a transport equation
\be \frac{df(\mathbf{x},\mathbf{v},t)}{dt} = C [f(\mathbf{x},\mathbf{v},t)]\ee
that is an integro-differential equation for the one-particle distribution function $f(\mathbf{x},\mathbf{v},t)$. This distribution function depends on time because we deal with a nonequilibrium gas. The explicit equation for our boson gas is called the Boltzmann-Uehling-Uhlenbeck (BUU) equation and reads
\be \label{BUU} \frac{\pa f}{\pa t} + \frac{\mathbf{p}}{E(p)} \mathbf{\nabla} f = \int d \overline{\sigma} d \mathbf{p}_1 v_{\textrm{rel}} \left[f' f'_1 \left( 1+ \frac{(2 \pi)^3}{N} f\right) \left( 1+ \frac{(2\pi)^3}{N}f_1\right) - f f_1 \left( 1+ \frac{(2 \pi)^3}{N} f'\right) \left( 1+ \frac{(2 \pi)^3}{N} f'_1\right) \right].\ee

The first step is a linearization of Eq.~(\ref{BUU}). Slightly out of equilibrium,
\be f \simeq  f_0 + \delta f= f_0 \left( 1 + \frac{\chi}{T} \right),\ee
where $f_0$ is the Bose-Einstein distribution function, solution to the transport equation with vanishing collision term. Following the Chapman-Enskog expansion, the perturbation function $\chi$ must be proportional to the gradients of the hydrodynamical fields out of equilibrium. For the viscosity these are nothing but the velocity fields. Following the lines of \cite{Dobado:2003wr} one can parametrize $\delta f$ as

\be f = f_0 \left[ 1+ \frac{g(p)}{T} \Delta_{ij} \tilde{V}^{ij} \right], \ee

where $\Delta_{ij} \equiv p_ip_j  -\frac{1}{3} \delta_{ij} p^2 $ and $\tilde{V}_{ij}$ represents the gradient of the spatial velocity field, $V_i$ :
\be \tilde{V}_{ij} = \frac{1}{2} (\pa_i V_j + \pa_j V_i) - \frac{1}{3} \pa_k V^k \delta_{ij}.\ee

The scalar function $g(p)$ depends only on moduli of  momenta and  is expanded in a convenient polynomial basis
\be \label{expansion} g(x)= \sum_{s=0}^{\infty} B_s P^s (x;y,z),\ee
where $x \equiv p^2/m^2_{\pi}$, $y \equiv m_{\pi} /T$ and $z$ is the fugacity of the boson gas.

This polynomial functions are used as variational {\it ansatz} to systematically solve the linearized equation.

The microscopic expression for the viscosity in terms of $g(p)$ becomes
\be \eta = -\frac{1}{10T} \ \int \frac{d\mathbf{p}}{E(p)} f_0 p_i p_j \Delta^{ij} g(p),\ee
where $g(p)$ should be determined by solving the BUU equation.

Taking only the first term in the expansion (\ref{expansion}) and inserting it into the viscosity we obtain
\be \eta = - \frac{2 \pi}{15 T} \frac{N m^6_{\pi}}{(2 \pi)^3} A^0_{5/2} (y,z) \ B_0,\ee

with the function
\be A^0_{5/2} (y,z)= \int_0^\infty dx \ \frac{x^{5/2}}{\sqrt{1+x} (z^{-1} e^{y (\sqrt{1+x}-1)}-1)} \ee

and $B_0=C_{\pi}/A_{11}^{\pi}$, where

\be C_{\pi} = \frac{4\pi m_{\pi}^6 N}{3 (2\pi)^3} A^0_{5/2} (y,z)\ee
and the collision integral

\be \label{collision} A^{\pi}_{11} = \frac{N^2}{z^2 (2\pi)^6} \int d\overline{\sigma} v_{\textrm{rel}} d \mathbf{p} d \mathbf{p_1} e^{\beta(E-2m_{\pi})} f_0 f_{01} f'_{0} f'_{01} \left(1-z e^{\beta(E-m_{\pi})} \right)  \left(\delta_{ik} \delta_{jl} - \frac{1}{3} \delta_{ij} \delta_{kl} \right) p^i p^j \ee
\be \label{intcol} \times  \left[ p'^kp'^l (1-e^{-\beta(E'-\mu)}) - p^k p^l (1-e^{-\beta(E-\mu)}) +p'^k_1 p'^l_1 (1-e^{-\beta(E_1'-\mu)}) - p^k_1 p^l_1 (1-e^{-\beta(E_1-\mu)}) \right].\ee

This collision integral requires intense computational work. Once the kinematics are analyzed and conserved quantities employed, the collision term turns into a complicated five-dimensional integral, handled with a Monte Carlo integration routine. The calculation has already been described in \cite{Dobado:2003wr}.

\section{Shear Viscosity in the Green-Kubo formalism}
\label{kubovisco}
In this Appendix we briefly review the computation of the shear viscosity in the Green--Kubo formalism, based largely on the works of \cite{FernandezFraile:2009mi, FernandezFraile:2007zz}. The Green--Kubo formulas for transport coefficients are based on the Linear Response Theory. If the response of a system to an external perturbation is considered to be linear in the perturbing field, the transport coefficient associated with a conserved quantity can be expressed as an expected value at equilibrium of the commutator of the corresponding (perturbed) Noether current.

For the shear viscosity,
\be \eta= \frac{1}{20} \lim_{\omega \rightarrow 0^+} \lim_{|\mathbf{p}|\rightarrow 0^+} \frac{1}{\omega} \int d^4 \mathbf{x} \ e^{i \mathbf{p} \cdot \mathbf{x}} \ \la \left[ \hat{\pi} ^{ij}(x), \hat{\pi}_{ij} (0)\right] \ra,\ee
where $\hat{\pi}^{ij}=\hat{T}^{ij}-g^{ij} \hat{T}^k_k/3$. Coincidence with the transport equation approach occurs at low temperature. In this limit, the simplest dominant resummation~\cite{FernandezFraile:2009mi} gives 
\be \eta^{(0)}=\frac{1}{10 \pi^2 T} \int_0 ^{\infty} d|\mathbf{p}| \frac{|\mathbf{p}|^6}{E_p^2 \ \Gamma(p)} n_B(E_p) [1+ n_B(E_p)],\ee
where $E_p^2=p^2+m_{\pi}^2$, $n_B(E)$ is the Bose-Einstein occupation number
\be n_B(E_p)=\frac{1}{e^{\beta E_p}-1}\ee

and $\Gamma(p)$ is the pion width in the thermal bath. The particle width can be understood as the inverse of the collision time in the gas and is related to the pion self-energy by
\be \Gamma(p)= - \frac{\textrm{Im } \Pi_R(E_p, \mathbf{p})}{2E_p}, \ee
where the retarded self-energy $\Pi_R(E_p,\mathbf{p})$ is given by the quantum fluctuations in the medium that change the dispersion relation of the pion.

In the Dilute Gas Approximation ($\beta E_p \gg 1$) the thermal width is given by
\be \Gamma(p) \simeq \frac{1}{2} \int \frac{d^3 k}{(2\pi)^3} \ \sigma_{\textrm{tot}}(s) \frac{\sqrt{s(s-4m_{\pi}^2)}}{2E_kE_p} \ e^{-E_k/T},\ee

where as usual $E_k=\sqrt{k^2+m_{\pi}^2} \rightarrow k$ in the chiral limit, $s(k,p)$ is the Mandelstam variable and $\sigma_{\textrm{tot}}(s)$ is the total pion-pion scattering cross section for the L$\sigma$M obtained in (\ref{sigmatot}). 

As in the kinetic theory approach the coefficient of shear viscosity turns out to be inversely proportional to the cross-section.

\section{Applicability of the transport equation}
\label{app:density}

The applicability of transport equations is rooted in Boltzmann's approximation of ``molecular chaos''. A rigorous formulation of quantum field-theory (multiparticle quantum mechanics) in a context appropriate to study statistical mechanics can well start with the Wigner function, and deduce from it the Bogoliubov-Born-Green-Kirkwood-Yvon hierarchy of equations for multiparticle distribution functions. Decoupling the lowest order equation for the one-particle distribution function $f({\bf x},{\bf{p}},t)$ from the rest is an approximation of low density; two successive collisions of the same quantum must be uncorrelated.

This condition is tantamount to stating that the mean free path (dependent on density and average cross section)
$$\lambda \propto \frac{1}{n(T)\bar{\sigma}}$$
is much smaller than the reach of the interaction, that can be expressed as the scattering length at low energies, or more generally as the square root of the averaged cross-section
\ba \nonumber
\frac{1}{n(T) \bar{\sigma}} >> \sqrt{\bar{\sigma}} \\
n(T) \bar{\sigma}^{3/2}<<1 \ .
\ea

\begin{figure}
\includegraphics[height=2.5in]{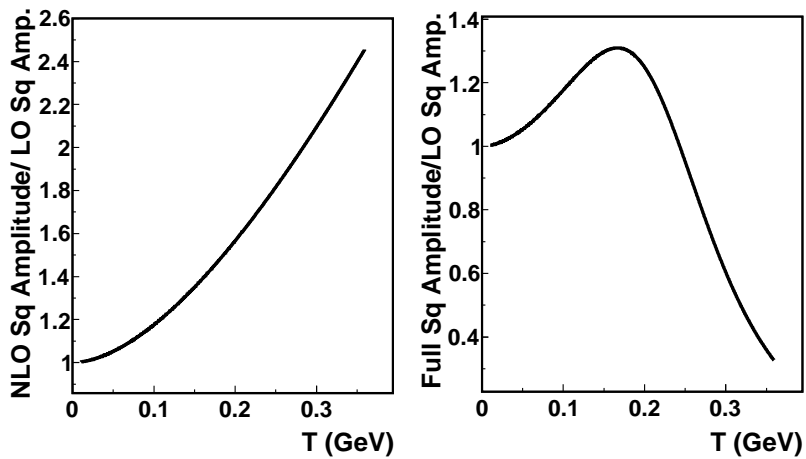}
\caption{Left panel: Ratio of NLO to LO terms in a derivative expansion of Eq.~(\ref{sqscamp}).
Right panel: Ratio of full to LO squared scattering amplitude.}
\label{fig:ratios}
\end{figure}

\begin{figure}
\includegraphics[height=2.5in]{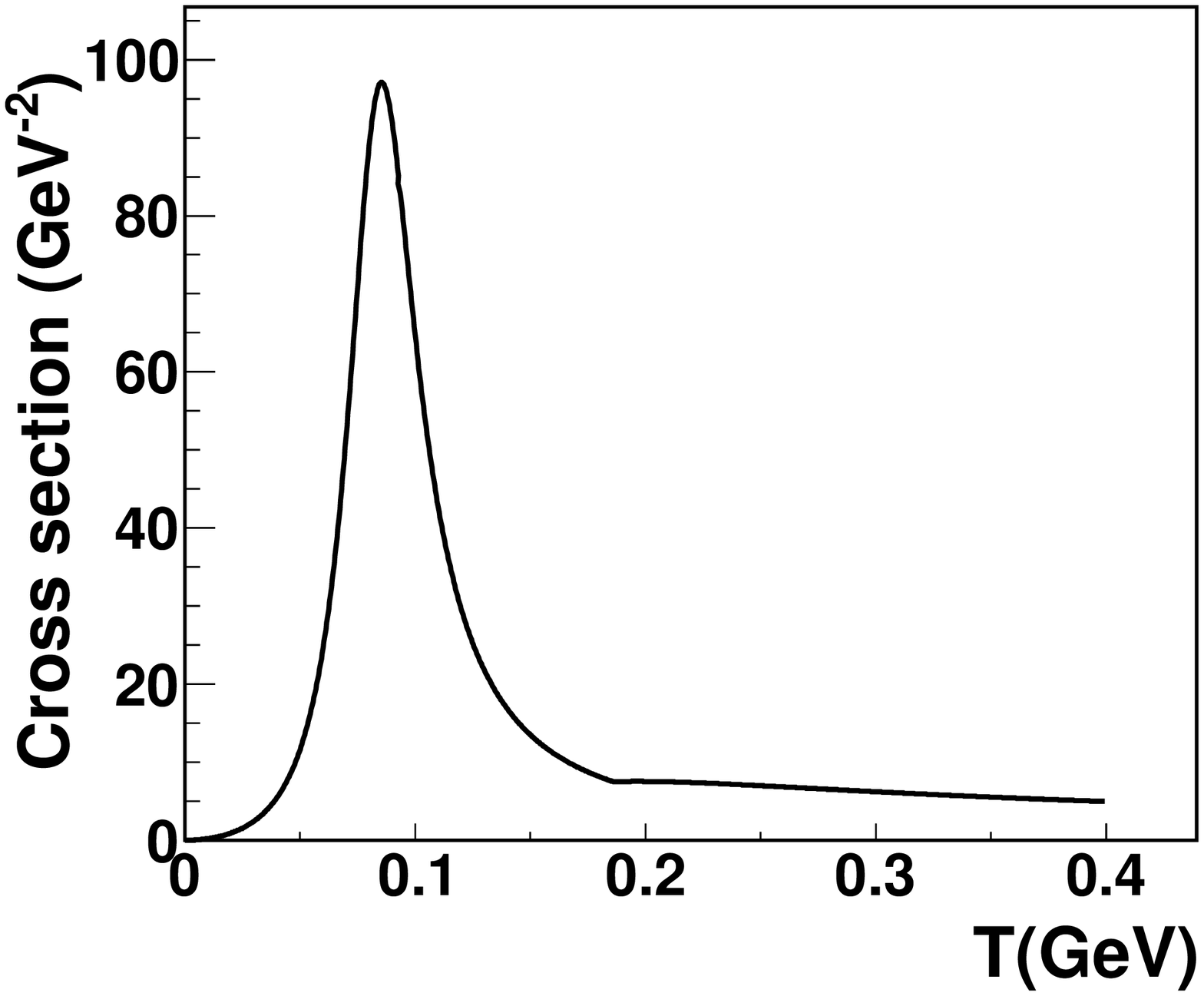}\hspace{1.3cm}
\includegraphics[height=2.5in]{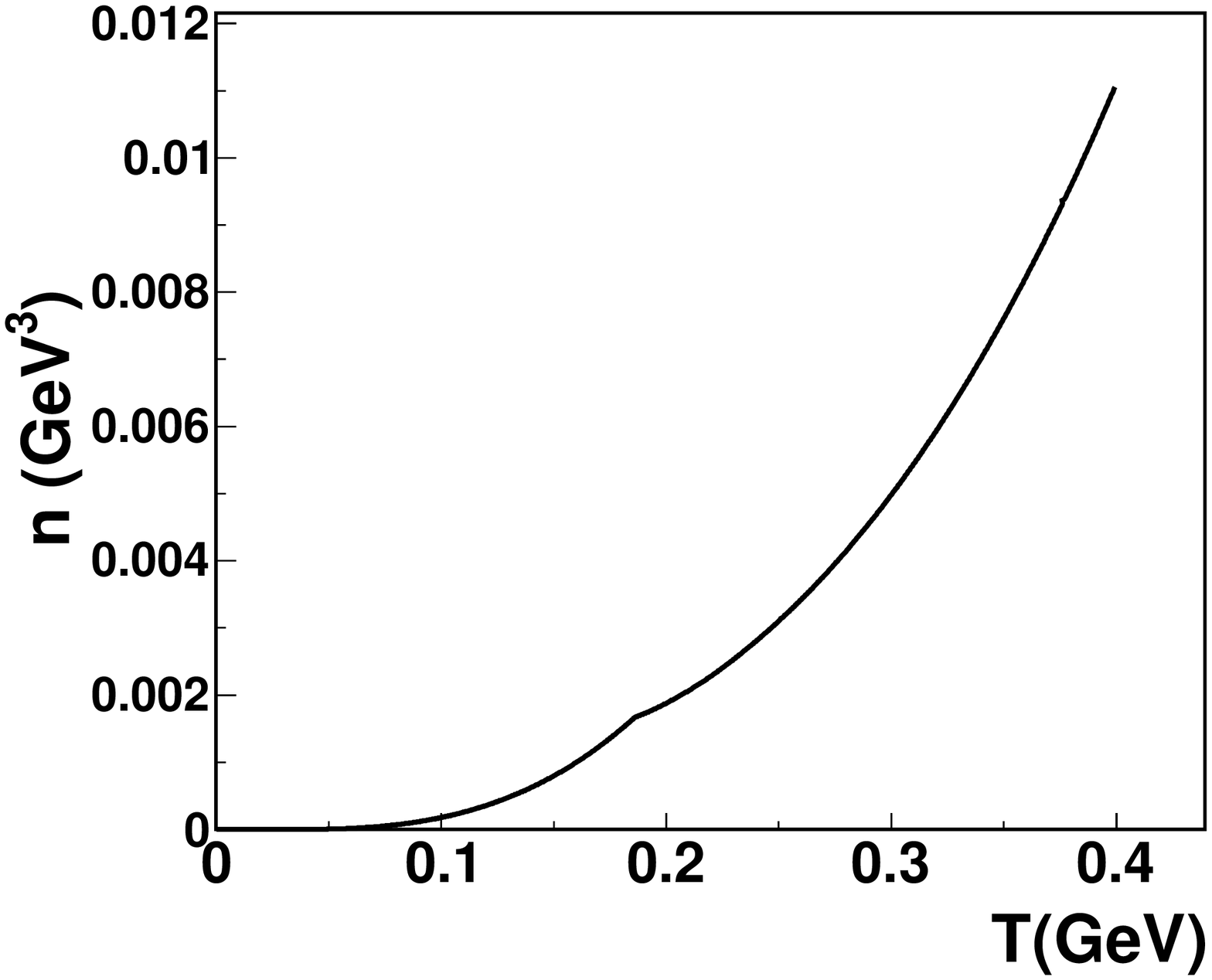}
\caption{Left panel: Typical cross section in the Linear Sigma model. We have taken an approximate average for Mandelstam $s$ given by $s\simeq 10 T^2$ at a typical temperature $T$ (with $m_{\pi} (0)=0$). 
Right panel: Particle number density for a free Bose gas in the same temperature range.}
\label{fig:crosssec}
\end{figure}

To satisfy this relation the interaction does not necessarily need to be very weak. As long as the free Bose density is a reasonable approximation (the system remains in a gaslike phase, as is the case for the L$\sigma$M), one can accept moderately strong interactions.
Simple criteria for the strength of the interaction are to examine the ratio of the leading order (LO) and next-to-leading order (NLO) derivative expansion of the squared scattering amplitude Eq.~(\ref{sqscamp})
\ba
\ar A_{\rm LO}\ar^2= \frac{s^2}{v^4}\\
\ar A_{\rm NLO}\ar^2= \frac{s^2}{v^4} \left[ 1+ \frac{2s}{M_R^2} - \frac{2sN^2}{32 \pi^2 v^2} \log \left( \frac{s}{\mu^2} \right) \right]
\ea
or the ratio of the LO squared amplitude to the total square amplitude. The two inverse ratios are plotted in Fig.~\ref{fig:ratios}
They clearly demonstrate that the interactions are indeed strong, thus requiring our large $N$ resummation. We have however seen in Sec.~\ref{sec:KSS} that the entropy density (and hence very likely the particle density) is close to the value in a free Bose gas.
In Fig. \ref{fig:crosssec} we then plot the maximum possible value of the total cross-section, reached at the peak of the sigma resonance
\be \label{analytic} 
\sigma_M(s) = \frac{s}{32 \pi (v^2)^2} \frac{1}{\left[ 1- \frac{s}{M_R^2} + \frac{Ns}{32 \pi^2 v^2} \log \left( \frac{s}{\mu^2}\right)\right]^2+ \left( \frac{sN}{32 \pi v^2} \right)^2}\ .
\ee
and this particle number density
\be
 n (T) = \int 4 \pi dp p^2 \frac{N/(2\pi)^3}{e^{\beta \sqrt{p^2+m_{\pi}^2(T)}}-1}\ .
\ee

\begin{figure}
\includegraphics[height=2.5in]{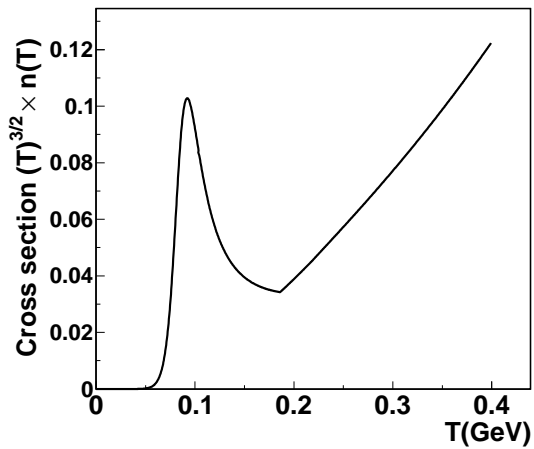}
\caption{The product $\bar{\sigma}^{3/2} \ n(T)$, significantly smaller than 1 as seen in this graph, shows that the Boltzmann-Uehling-Uhlenbeck equation is applicable in the range of temperatures that we have studied.}
\label{fig:sigmadens}
\end{figure}

Finally we construct the product $n(T) \bar{\sigma}^{3/2}$ and plot it in Fig.~\ref{fig:sigmadens}. The value of this parameter is seen to be of order 0.1 in our entire temperature range. Its actual value will be even smaller with an averaged cross-section (we are using its maximum possible value at a given temperature). Therefore there is no reason to question the validity of Boltzmann's equation in the model.

\section{Saturation of the pole mass}
\label{app:polemass}

\begin{figure}
\includegraphics[height=2.2in]{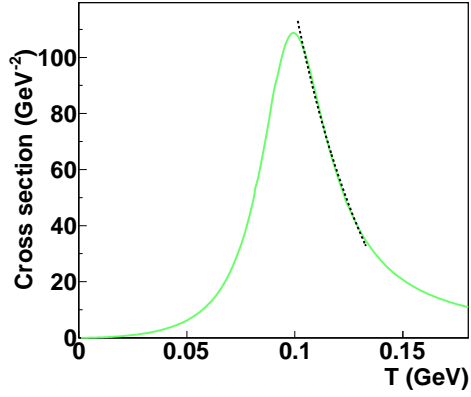}
\caption{The trailing tail of the $\sigma$ resonance occurs just before the phase transition temperature $T_c$. Because the cross-section drops rapidly, the viscosity raises rapidly (faster than the entropy density $s\propto T^3)$ and $\eta/s$ grows slightly before the phase transition.
}
\label{fig:crosssec2}
\end{figure}

To further understand why the minimum of $\eta/s$ occurs just before the phase transition, we now perform a scaling analysis. For this, let us note that dimensionally and as a very rough average (not comparable with the quality calculations presented in the main text) the scaling laws are
\be \eta \sim \frac{T}{\sigma}; \qquad s \sim T^3 .\ee
From the cross section in Fig.~\ref{fig:crosssec2} one can see that the maximum of the $\pi \pi$ cross section (sigma pole) occurs somewhat before the phase transition. 
The trailing tail of the resonance is well fit (see figure) by a power-law
\be 
\sigma = A^{2+k} \frac{1}{T^k}\ , 
\ee
where $A$ is a constant with [mass dimension] = 1. Then, just before the minimum:
\be \eta/s \sim T^{k-2}.\ee
The fit  gives the value $k = 2.527 \pm 0.020 $ that means that, for temperatures just above its minimum, 
$\eta/s \sim T^{0.527}$; that is, the KSS coefficient begins to grow somewhat slowly toward the phase transition.
The same conclusion can be reached by analytically examining the scattering amplitude 
in Eq.~(\ref{sqscamp}).

Next we examine the connection of this turning of the cross section with the order parameter, the sigma condensate.
\begin{figure}[h]
\includegraphics[height=2.5in]{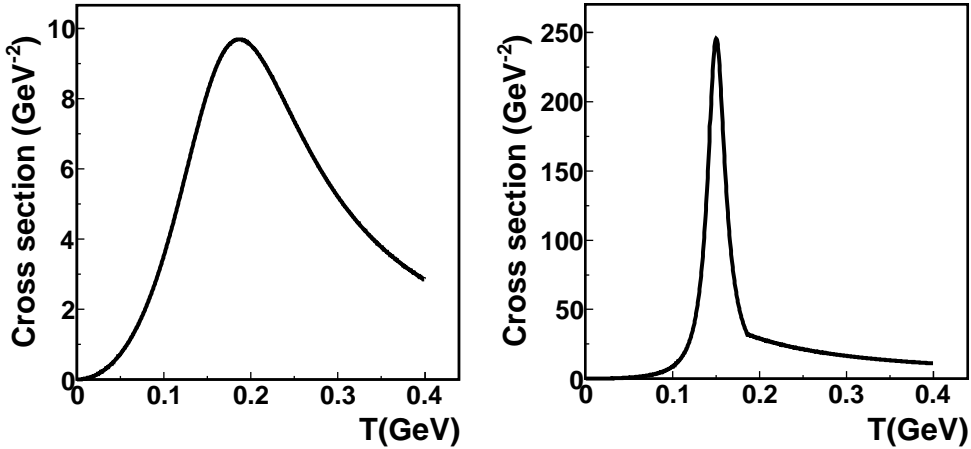}
\caption{$\pi\pi$ cross section at finite temperature in the L$\sigma$M with (right) and without (left) the variation of the order parameter $\la \sigma \ra(T)$ that affects the amplitude in Eq.~(\ref{sqscamp}) through $F(T)$. 
}
\label{fig:crosssec3}
\end{figure}
In Fig.~\ref{fig:crosssec3} we plot the cross section that we have just commented on and the cross section where we have artificially removed the variation of the order parameter $\la \sigma (T) \ra$, and fixed it as a constant. As can be seen, the variation of the order parameter with $T$ makes the resonance narrower (and only slightly shifted toward smaller energies). This plotted behavior can be followed analytically from Eq.~(\ref{sqscamp}).
Therefore, the slight rise of $\eta/s$ with temperature before the phase transition
is indeed a direct consequence of the decrease of the order parameter. 

\begin{figure}
\includegraphics[height=2.6in]{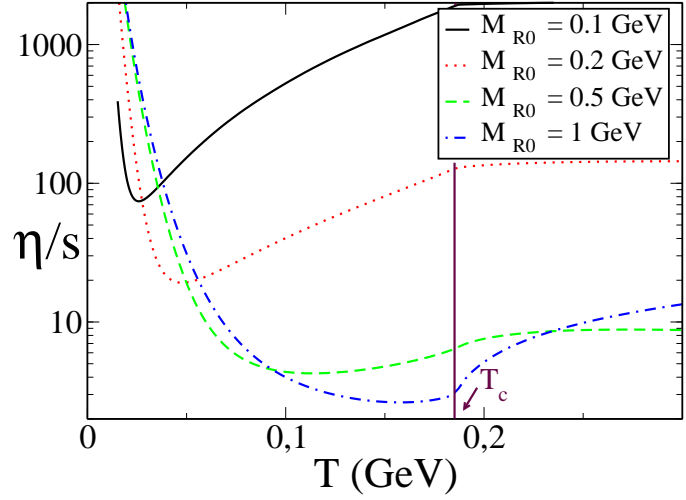}
\caption{$\eta/s$ and its dependence on the parameter $M_{R0}$, the tree-level (not the pole mass) $\sigma$ mass at T=0. The minimum of $\eta/s$ increases with $M_{R0}$, but does not exceed $T_c$.}
\label{fig:MRdependence}
\end{figure}

The pole mass position is highly dependent on $M_{R0}$, a parameter controlling the propagator at tree level. Both the maximum of the cross section and the minimum of $\eta/s$ move with $M_{R0}$. Just as in Fig. \ref{fig:Fdependence} we plot the dependence of $\eta/s$ with $M_{R0}$ in Fig. \ref{fig:MRdependence}.

\begin{figure}
\includegraphics[height=2.6in]{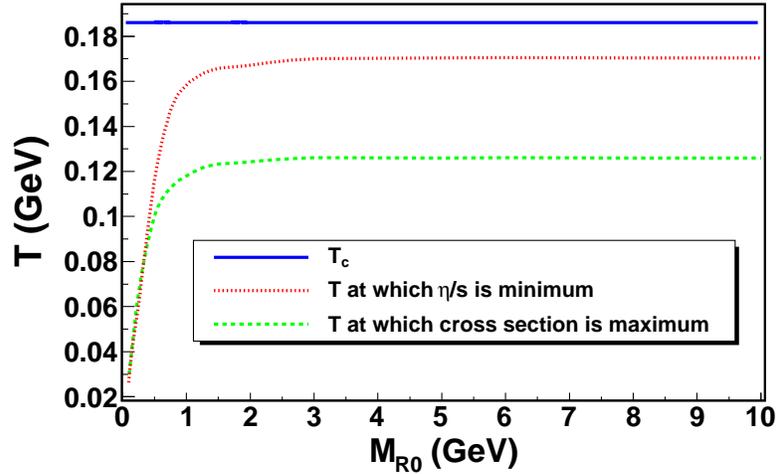}
\caption{Saturation of the temperature at which the cross section is maximum and the temperature at which $\eta/s$ is minimum. Both are related because $\eta/s \propto 1/\sigma$ as shown in both Boltzmann and Green-Kubo formalisms. The critical temperature is always above these temperatures for any value of $M_{R0}$.}
\label{fig:saturation}
\end{figure}

In this figure we show that the exact temperature of the minimum grows with $M_{R0}$. From this fact, one could ask if there exists a value of $M_{R0}$ for which the minimum of $\eta/s$ is located at $T_c$ or above it. However our analysis stands as presented; increasing the value of $M_{R0}$ we observe that the pole mass in the squared amplitude saturates. The minimum of $\eta/s$ saturates as well when $M_{R0} \rightarrow \infty$. In Figure \ref{fig:saturation} we plot the temperatures at which the maximum of the cross section is located, and the one at which the minimum of $\eta/s$ is reached. Both values tend to a constant always below $T_c$ when $M_{R0}$ is arbitrarily increased. We conclude that the minimum of $\eta/s$ takes place always below $T_c$ and never at or above it.

\bibliographystyle{apsrev}
\bibliography{DLLETR_LSM2}

\begin{thebibliography}{42}
\expandafter\ifx\csname natexlab\endcsname\relax\def\natexlab#1{#1}\fi
\expandafter\ifx\csname bibnamefont\endcsname\relax
  \def\bibnamefont#1{#1}\fi
\expandafter\ifx\csname bibfnamefont\endcsname\relax
  \def\bibfnamefont#1{#1}\fi
\expandafter\ifx\csname citenamefont\endcsname\relax
  \def\citenamefont#1{#1}\fi
\expandafter\ifx\csname url\endcsname\relax
  \def\url#1{\texttt{#1}}\fi
\expandafter\ifx\csname urlprefix\endcsname\relax\def\urlprefix{URL }\fi
\providecommand{\bibinfo}[2]{#2}
\providecommand{\eprint}[2][]{\url{#2}}

\bibitem[{\citenamefont{Kovtun et~al.}(2005)\citenamefont{Kovtun, Son, and
  Starinets}}]{Kovtun:2004de}
\bibinfo{author}{\bibfnamefont{P.}~\bibnamefont{Kovtun}},
  \bibinfo{author}{\bibfnamefont{D.~T.} \bibnamefont{Son}}, \bibnamefont{and}
  \bibinfo{author}{\bibfnamefont{A.~O.} \bibnamefont{Starinets}},
  \bibinfo{journal}{Phys. Rev. Lett.} \textbf{\bibinfo{volume}{94}},
  \bibinfo{pages}{111601} (\bibinfo{year}{2005}), \eprint{hep-th/0405231}.

\bibitem[{\citenamefont{Dobado et~al.}(2009{\natexlab{a}})\citenamefont{Dobado,
  Llanes-Estrada, and Torres-Rincon}}]{Dobado:2008vt}
\bibinfo{author}{\bibfnamefont{A.}~\bibnamefont{Dobado}},
  \bibinfo{author}{\bibfnamefont{F.~J.} \bibnamefont{Llanes-Estrada}},
  \bibnamefont{and} \bibinfo{author}{\bibfnamefont{J.~M.}
  \bibnamefont{Torres-Rincon}}, \bibinfo{journal}{Phys. Rev.}
  \textbf{\bibinfo{volume}{D79}}, \bibinfo{pages}{014002}
  (\bibinfo{year}{2009}{\natexlab{a}}), \eprint{hep-ph/0803.3275}.

\bibitem[{\citenamefont{Mueller et~al.}(2009)\citenamefont{Mueller, Schmalian,
  and Fritz}}]{mueller-2009}
\bibinfo{author}{\bibfnamefont{M.}~\bibnamefont{Mueller}},
  \bibinfo{author}{\bibfnamefont{J.}~\bibnamefont{Schmalian}},
  \bibnamefont{and} \bibinfo{author}{\bibfnamefont{L.}~\bibnamefont{Fritz}},
  \bibinfo{journal}{Phys. Rev. Lett.} \textbf{\bibinfo{volume}{103}},
  \bibinfo{pages}{025301} (\bibinfo{year}{2009}).

\bibitem[{\citenamefont{Schaefer}(2007)}]{schaefer-2007-76}
\bibinfo{author}{\bibfnamefont{T.}~\bibnamefont{Schaefer}},
  \bibinfo{journal}{Phys. Rev. A} \textbf{\bibinfo{volume}{76}},
  \bibinfo{pages}{063618} (\bibinfo{year}{2007}).

\bibitem[{\citenamefont{Turlapov et~al.}(2008)\citenamefont{Turlapov, Kinast,
  Clancy, Luo, Joseph, and Thomas}}]{Turlapovetal}
\bibinfo{author}{\bibfnamefont{A.}~\bibnamefont{Turlapov}},
  \bibinfo{author}{\bibfnamefont{J.}~\bibnamefont{Kinast}},
  \bibinfo{author}{\bibfnamefont{B.}~\bibnamefont{Clancy}},
  \bibinfo{author}{\bibfnamefont{L.}~\bibnamefont{Luo}},
  \bibinfo{author}{\bibfnamefont{J.}~\bibnamefont{Joseph}}, \bibnamefont{and}
  \bibinfo{author}{\bibfnamefont{J.}~\bibnamefont{Thomas}},
  \bibinfo{journal}{J. Low Temp. Phys.} \textbf{\bibinfo{volume}{150}},
  \bibinfo{pages}{567} (\bibinfo{year}{2008}).

\bibitem[{\citenamefont{Csernai et~al.}(2006)\citenamefont{Csernai, Kapusta,
  and McLerran}}]{Csernai:2006zz}
\bibinfo{author}{\bibfnamefont{L.~P.} \bibnamefont{Csernai}},
  \bibinfo{author}{\bibfnamefont{J.~I.} \bibnamefont{Kapusta}},
  \bibnamefont{and} \bibinfo{author}{\bibfnamefont{L.~D.}
  \bibnamefont{McLerran}}, \bibinfo{journal}{Phys. Rev. Lett.}
  \textbf{\bibinfo{volume}{97}}, \bibinfo{pages}{152303}
  (\bibinfo{year}{2006}), \eprint{nucl-th/0604032}.

\bibitem[{\citenamefont{Gavin and Abdel-Aziz}(2006)}]{Gavin:2006xd}
\bibinfo{author}{\bibfnamefont{S.}~\bibnamefont{Gavin}} \bibnamefont{and}
  \bibinfo{author}{\bibfnamefont{M.}~\bibnamefont{Abdel-Aziz}},
  \bibinfo{journal}{Phys. Rev. Lett.} \textbf{\bibinfo{volume}{97}},
  \bibinfo{pages}{162302} (\bibinfo{year}{2006}), \eprint{nucl-th/0606061}.

\bibitem[{\citenamefont{Collaboration}(2009)}]{Collaboration:2009yd}
\bibinfo{author}{\bibfnamefont{C.}~\bibnamefont{Collaboration}}
  (\bibinfo{year}{2009}), \eprint{nucl-ex/0907.2799}.

\bibitem[{\citenamefont{Demir and Bass}(2009)}]{Demir:2008tr}
\bibinfo{author}{\bibfnamefont{N.}~\bibnamefont{Demir}} \bibnamefont{and}
  \bibinfo{author}{\bibfnamefont{S.~A.} \bibnamefont{Bass}},
  \bibinfo{journal}{Phys. Rev. Lett.} \textbf{\bibinfo{volume}{102}},
  \bibinfo{pages}{172302} (\bibinfo{year}{2009}), \eprint{nucl-th/0812.2422}.

\bibitem[{\citenamefont{Danielewicz and Gyulassy}(1985)}]{Danielewicz:1984ww}
\bibinfo{author}{\bibfnamefont{P.}~\bibnamefont{Danielewicz}} \bibnamefont{and}
  \bibinfo{author}{\bibfnamefont{M.}~\bibnamefont{Gyulassy}},
  \bibinfo{journal}{Phys. Rev.} \textbf{\bibinfo{volume}{D31}},
  \bibinfo{pages}{53} (\bibinfo{year}{1985}).

\bibitem[{\citenamefont{Buchel et~al.}(2009)\citenamefont{Buchel, Myers, and
  Sinha}}]{Buchel:2008vz}
\bibinfo{author}{\bibfnamefont{A.}~\bibnamefont{Buchel}},
  \bibinfo{author}{\bibfnamefont{R.~C.} \bibnamefont{Myers}}, \bibnamefont{and}
  \bibinfo{author}{\bibfnamefont{A.}~\bibnamefont{Sinha}},
  \bibinfo{journal}{JHEP} \textbf{\bibinfo{volume}{03}}, \bibinfo{pages}{084}
  (\bibinfo{year}{2009}), \eprint{hep-th/0812.2521}.

\bibitem[{\citenamefont{Sinha and Myers}(2009)}]{Sinha:2009ev}
\bibinfo{author}{\bibfnamefont{A.}~\bibnamefont{Sinha}} \bibnamefont{and}
  \bibinfo{author}{\bibfnamefont{R.~C.} \bibnamefont{Myers}}
  (\bibinfo{year}{2009}), \eprint{hep-th/0907.4798}.

\bibitem[{\citenamefont{Dobado et~al.}(2009{\natexlab{b}})\citenamefont{Dobado,
  Llanes-Estrada, and Torres-Rincon}}]{Dobado:2008jr}
\bibinfo{author}{\bibfnamefont{A.}~\bibnamefont{Dobado}},
  \bibinfo{author}{\bibfnamefont{F.~J.} \bibnamefont{Llanes-Estrada}},
  \bibnamefont{and} \bibinfo{author}{\bibfnamefont{J.~M.}
  \bibnamefont{Torres-Rincon}}, \bibinfo{journal}{AIP Conf. Proc.}
  \textbf{\bibinfo{volume}{1116}}, \bibinfo{pages}{421}
  (\bibinfo{year}{2009}{\natexlab{b}}), \eprint{hep-ph/0812.2203}.

\bibitem[{\citenamefont{Chen et~al.}(2008)\citenamefont{Chen, Huang, Li,
  Nakano, and Yang}}]{Chen:2007jq}
\bibinfo{author}{\bibfnamefont{J.-W.} \bibnamefont{Chen}},
  \bibinfo{author}{\bibfnamefont{M.}~\bibnamefont{Huang}},
  \bibinfo{author}{\bibfnamefont{Y.-H.} \bibnamefont{Li}},
  \bibinfo{author}{\bibfnamefont{E.}~\bibnamefont{Nakano}}, \bibnamefont{and}
  \bibinfo{author}{\bibfnamefont{D.-L.} \bibnamefont{Yang}},
  \bibinfo{journal}{Phys. Lett.} \textbf{\bibinfo{volume}{B670}},
  \bibinfo{pages}{18} (\bibinfo{year}{2008}), \eprint{hep-ph/0709.3434}.

\bibitem[{\citenamefont{Sasaki and Redlich}(2008)}]{Sasaki:2008um}
\bibinfo{author}{\bibfnamefont{C.}~\bibnamefont{Sasaki}} \bibnamefont{and}
  \bibinfo{author}{\bibfnamefont{K.}~\bibnamefont{Redlich}}
  (\bibinfo{year}{2008}), \eprint{hep-ph/0811.4708}.

\bibitem[{\citenamefont{Teaney}(2003)}]{Teaney:2003kp}
\bibinfo{author}{\bibfnamefont{D.}~\bibnamefont{Teaney}},
  \bibinfo{journal}{Phys. Rev.} \textbf{\bibinfo{volume}{C68}},
  \bibinfo{pages}{034913} (\bibinfo{year}{2003}), \eprint{nucl-th/0301099}.

\bibitem[{\citenamefont{Arnold et~al.}(2003)\citenamefont{Arnold, Moore, and
  Yaffe}}]{Arnold:2003zc}
\bibinfo{author}{\bibfnamefont{P.}~\bibnamefont{Arnold}},
  \bibinfo{author}{\bibfnamefont{G.~D.} \bibnamefont{Moore}}, \bibnamefont{and}
  \bibinfo{author}{\bibfnamefont{L.~G.} \bibnamefont{Yaffe}},
  \bibinfo{journal}{JHEP} \textbf{\bibinfo{volume}{05}}, \bibinfo{pages}{051}
  (\bibinfo{year}{2003}), \eprint{hep-ph/0302165}.

\bibitem[{\citenamefont{Noronha-Hostler
  et~al.}(2008)\citenamefont{Noronha-Hostler, Noronha, and Greiner}}]{Noronha}
\bibinfo{author}{\bibfnamefont{J.}~\bibnamefont{Noronha-Hostler}},
  \bibinfo{author}{\bibfnamefont{J.}~\bibnamefont{Noronha}}, \bibnamefont{and}
  \bibinfo{author}{\bibfnamefont{C.}~\bibnamefont{Greiner}}
  (\bibinfo{year}{2008}), \eprint{nucl-th/0811.1571}.

\bibitem[{\citenamefont{Aarts and Martinez~Resco}(2005)}]{Aarts:2005vc}
\bibinfo{author}{\bibfnamefont{G.}~\bibnamefont{Aarts}} \bibnamefont{and}
  \bibinfo{author}{\bibfnamefont{J.~M.} \bibnamefont{Martinez~Resco}},
  \bibinfo{journal}{JHEP} \textbf{\bibinfo{volume}{03}}, \bibinfo{pages}{074}
  (\bibinfo{year}{2005}), \eprint{hep-ph/0503161}.

\bibitem[{\citenamefont{Coleman et~al.}(1974)\citenamefont{Coleman, Jackiw, and
  Politzer}}]{Coleman:1974jh}
\bibinfo{author}{\bibfnamefont{S.~R.} \bibnamefont{Coleman}},
  \bibinfo{author}{\bibfnamefont{R.}~\bibnamefont{Jackiw}}, \bibnamefont{and}
  \bibinfo{author}{\bibfnamefont{H.~D.} \bibnamefont{Politzer}},
  \bibinfo{journal}{Phys. Rev.} \textbf{\bibinfo{volume}{D10}},
  \bibinfo{pages}{2491} (\bibinfo{year}{1974}).

\bibitem[{\citenamefont{Dobado and Morales}(1995)}]{Dobado:1994fd}
\bibinfo{author}{\bibfnamefont{A.}~\bibnamefont{Dobado}} \bibnamefont{and}
  \bibinfo{author}{\bibfnamefont{J.}~\bibnamefont{Morales}},
  \bibinfo{journal}{Phys. Rev.} \textbf{\bibinfo{volume}{D52}},
  \bibinfo{pages}{2878} (\bibinfo{year}{1995}), \eprint{hep-ph/9407321}.

\bibitem[{\citenamefont{Dobado et~al.}(1997)\citenamefont{Dobado, Gomez-Nicola,
  Maroto, and Pelaez}}]{Dobado:1997jx}
\bibinfo{author}{\bibfnamefont{A.}~\bibnamefont{Dobado}},
  \bibinfo{author}{\bibfnamefont{A.}~\bibnamefont{Gomez-Nicola}},
  \bibinfo{author}{\bibfnamefont{A.~L.} \bibnamefont{Maroto}},
  \bibnamefont{and} \bibinfo{author}{\bibfnamefont{J.~R.} \bibnamefont{Pelaez}}
  (\bibinfo{year}{1997}), \bibinfo{note}{{N.Y., Springer-Verlag, 1997. (Texts
  and Monographs in Physics)}}.

\bibitem[{\citenamefont{Petropoulos}(2004)}]{Petropoulos:2004bt}
\bibinfo{author}{\bibfnamefont{N.}~\bibnamefont{Petropoulos}}
  (\bibinfo{year}{2004}), \eprint{hep-ph/0402136}.

\bibitem[{\citenamefont{Aarts and Martinez~Resco}(2003)}]{Aarts:2003bk}
\bibinfo{author}{\bibfnamefont{G.}~\bibnamefont{Aarts}} \bibnamefont{and}
  \bibinfo{author}{\bibfnamefont{J.~M.} \bibnamefont{Martinez~Resco}},
  \bibinfo{journal}{Phys. Rev.} \textbf{\bibinfo{volume}{D68}},
  \bibinfo{pages}{085009} (\bibinfo{year}{2003}), \eprint{hep-ph/0303216}.

\bibitem[{\citenamefont{Aarts and Martinez~Resco}(2004)}]{Aarts:2004sd}
\bibinfo{author}{\bibfnamefont{G.}~\bibnamefont{Aarts}} \bibnamefont{and}
  \bibinfo{author}{\bibfnamefont{J.~M.} \bibnamefont{Martinez~Resco}},
  \bibinfo{journal}{JHEP} \textbf{\bibinfo{volume}{02}}, \bibinfo{pages}{061}
  (\bibinfo{year}{2004}), \eprint{hep-ph/0402192}.

\bibitem[{\citenamefont{Cornwall et~al.}(1974)\citenamefont{Cornwall, Jackiw,
  and Tomboulis}}]{Cornwall:1974vz}
\bibinfo{author}{\bibfnamefont{J.~M.} \bibnamefont{Cornwall}},
  \bibinfo{author}{\bibfnamefont{R.}~\bibnamefont{Jackiw}}, \bibnamefont{and}
  \bibinfo{author}{\bibfnamefont{E.}~\bibnamefont{Tomboulis}},
  \bibinfo{journal}{Phys. Rev.} \textbf{\bibinfo{volume}{D10}},
  \bibinfo{pages}{2428} (\bibinfo{year}{1974}).

\bibitem[{\citenamefont{Weinberg and Wu}(1987)}]{Weinberg:1987vp}
\bibinfo{author}{\bibfnamefont{E.~J.} \bibnamefont{Weinberg}} \bibnamefont{and}
  \bibinfo{author}{\bibfnamefont{A.-q.} \bibnamefont{Wu}},
  \bibinfo{journal}{Phys. Rev.} \textbf{\bibinfo{volume}{D36}},
  \bibinfo{pages}{2474} (\bibinfo{year}{1987}).

\bibitem[{\citenamefont{Dobado and Llanes-Estrada}(2004)}]{Dobado:2003wr}
\bibinfo{author}{\bibfnamefont{A.}~\bibnamefont{Dobado}} \bibnamefont{and}
  \bibinfo{author}{\bibfnamefont{F.~J.} \bibnamefont{Llanes-Estrada}},
  \bibinfo{journal}{Phys. Rev.} \textbf{\bibinfo{volume}{D69}},
  \bibinfo{pages}{116004} (\bibinfo{year}{2004}), \eprint{hep-ph/0309324}.

\bibitem[{\citenamefont{Dobado et~al.}(2007)\citenamefont{Dobado,
  Llanes-Estrada, and Torres~Rincon}}]{Dobado:2007cv}
\bibinfo{author}{\bibfnamefont{A.}~\bibnamefont{Dobado}},
  \bibinfo{author}{\bibfnamefont{F.~J.} \bibnamefont{Llanes-Estrada}},
  \bibnamefont{and} \bibinfo{author}{\bibfnamefont{J.~M.}
  \bibnamefont{Torres~Rincon}} (\bibinfo{year}{2007}), \eprint{hep-ph/0702130}.

\bibitem[{\citenamefont{Jeon and Yaffe}(1996)}]{Jeon:1995zm}
\bibinfo{author}{\bibfnamefont{S.}~\bibnamefont{Jeon}} \bibnamefont{and}
  \bibinfo{author}{\bibfnamefont{L.~G.} \bibnamefont{Yaffe}},
  \bibinfo{journal}{Phys. Rev.} \textbf{\bibinfo{volume}{D53}},
  \bibinfo{pages}{5799} (\bibinfo{year}{1996}), \eprint{hep-ph/9512263}.

\bibitem[{\citenamefont{Hohenberg and Halperin}(1977)}]{Hohenberg:1977ym}
\bibinfo{author}{\bibfnamefont{P.~C.} \bibnamefont{Hohenberg}}
  \bibnamefont{and} \bibinfo{author}{\bibfnamefont{B.~I.}
  \bibnamefont{Halperin}}, \bibinfo{journal}{Rev. Mod. Phys.}
  \textbf{\bibinfo{volume}{49}}, \bibinfo{pages}{435} (\bibinfo{year}{1977}).

\bibitem[{\citenamefont{Son and Stephanov}(2004)}]{Son:2004iv}
\bibinfo{author}{\bibfnamefont{D.~T.} \bibnamefont{Son}} \bibnamefont{and}
  \bibinfo{author}{\bibfnamefont{M.~A.} \bibnamefont{Stephanov}},
  \bibinfo{journal}{Phys. Rev.} \textbf{\bibinfo{volume}{D70}},
  \bibinfo{pages}{056001} (\bibinfo{year}{2004}), \eprint{hep-ph/0401052}.

\bibitem[{\citenamefont{Moore and Saremi}(2008)}]{Moore:2008ws}
\bibinfo{author}{\bibfnamefont{G.~D.} \bibnamefont{Moore}} \bibnamefont{and}
  \bibinfo{author}{\bibfnamefont{O.}~\bibnamefont{Saremi}},
  \bibinfo{journal}{JHEP} \textbf{\bibinfo{volume}{09}}, \bibinfo{pages}{015}
  (\bibinfo{year}{2008}), \eprint{hep-ph/0805.4201}.

\bibitem[{\citenamefont{Sengers}(1985)}]{Senders:1985}
\bibinfo{author}{\bibfnamefont{J.~V.} \bibnamefont{Sengers}},
  \bibinfo{journal}{Int. J. Thermophys.} \textbf{\bibinfo{volume}{6}},
  \bibinfo{pages}{203} (\bibinfo{year}{1985}).

\bibitem[{\citenamefont{Antonov}(2009)}]{Antonov:2009xn}
\bibinfo{author}{\bibfnamefont{D.}~\bibnamefont{Antonov}}
  (\bibinfo{year}{2009}), \eprint{hep-ph/0905.3329}.

\bibitem[{\citenamefont{Hidaka and Pisarski}(2008)}]{Hidaka:2008dr}
\bibinfo{author}{\bibfnamefont{Y.}~\bibnamefont{Hidaka}} \bibnamefont{and}
  \bibinfo{author}{\bibfnamefont{R.~D.} \bibnamefont{Pisarski}},
  \bibinfo{journal}{Phys. Rev.} \textbf{\bibinfo{volume}{D78}},
  \bibinfo{pages}{071501} (\bibinfo{year}{2008}), \eprint{hep-ph/0803.0453}.

\bibitem[{\citenamefont{Gasser and Leutwyler}(1987)}]{Gasser:1986vb}
\bibinfo{author}{\bibfnamefont{J.}~\bibnamefont{Gasser}} \bibnamefont{and}
  \bibinfo{author}{\bibfnamefont{H.}~\bibnamefont{Leutwyler}},
  \bibinfo{journal}{Phys. Lett.} \textbf{\bibinfo{volume}{B184}},
  \bibinfo{pages}{83} (\bibinfo{year}{1987}).

\bibitem[{\citenamefont{Grayer et~al.}(1974)}]{Grayer:1974cr}
\bibinfo{author}{\bibfnamefont{G.}~\bibnamefont{Grayer}} \bibnamefont{et~al.},
  \bibinfo{journal}{Nucl. Phys.} \textbf{\bibinfo{volume}{B75}},
  \bibinfo{pages}{189} (\bibinfo{year}{1974}).

\bibitem[{\citenamefont{Rosselet et~al.}(1977)}]{Rosselet:1976pu}
\bibinfo{author}{\bibfnamefont{L.}~\bibnamefont{Rosselet}}
  \bibnamefont{et~al.}, \bibinfo{journal}{Phys. Rev.}
  \textbf{\bibinfo{volume}{D15}}, \bibinfo{pages}{574} (\bibinfo{year}{1977}).

\bibitem[{\citenamefont{Protopopescu et~al.}(1973)}]{Protopopescu:1973sh}
\bibinfo{author}{\bibfnamefont{S.~D.} \bibnamefont{Protopopescu}}
  \bibnamefont{et~al.}, \bibinfo{journal}{Phys. Rev.}
  \textbf{\bibinfo{volume}{D7}}, \bibinfo{pages}{1279} (\bibinfo{year}{1973}).

\bibitem[{\citenamefont{Fernandez-Fraile and
  Gomez~Nicola}(2009)}]{FernandezFraile:2009mi}
\bibinfo{author}{\bibfnamefont{D.}~\bibnamefont{Fernandez-Fraile}}
  \bibnamefont{and}
  \bibinfo{author}{\bibfnamefont{A.}~\bibnamefont{Gomez~Nicola}},
  \bibinfo{journal}{Eur. Phys. J.} \textbf{\bibinfo{volume}{C62}},
  \bibinfo{pages}{37} (\bibinfo{year}{2009}), \eprint{hep-ph/0902.4829}.

\bibitem[{\citenamefont{Fernandez-Fraile and
  Nicola}(2007)}]{FernandezFraile:2007zz}
\bibinfo{author}{\bibfnamefont{D.}~\bibnamefont{Fernandez-Fraile}}
  \bibnamefont{and} \bibinfo{author}{\bibfnamefont{A.~G.}
  \bibnamefont{Nicola}}, \bibinfo{journal}{Eur. Phys. J.}
  \textbf{\bibinfo{volume}{A31}}, \bibinfo{pages}{848} (\bibinfo{year}{2007}),
  \eprint{hep-ph/0610197}.

\end{thebibliography}

\end{document}